\documentclass[11pt]{article}
\pdfoutput=1

\usepackage[usenames]{xcolor}
\usepackage[english]{babel}
\usepackage[small]{caption}
\usepackage{verbatim}
\usepackage{multirow}
\usepackage{relsize}
\usepackage{amsmath,amssymb,amsbsy,amstext,amsfonts}
\usepackage{epsfig,multicol,bbm,simplewick}
\usepackage{mathtools}
\usepackage{enumitem}
\usepackage{csquotes}
\usepackage{tensor}
\usepackage{fix-cm}
\usepackage{booktabs}
\usepackage{graphicx}
\usepackage{bm}
\usepackage{subfigure}
\usepackage{float}
\usepackage[explicit]{titlesec}
\usepackage{enumitem}
\usepackage{setspace}
\usepackage{blindtext}
\usepackage{textpos}
\usepackage{etoolbox}
\usepackage[linktocpage=true]{hyperref}
\usepackage{anyfontsize}
\usepackage{listings,xfp}
\usepackage[dotinlabels]{titletoc}

\dottedcontents{section}[1.2em]{}{1.2em}{6pt}

\definecolor{jrcolor}{rgb}{.8,.1,.1}
 
\definecolor{jrcolor2}{rgb}{.6,.3,.3}

\usepackage[maxbibnames=99,style=numeric-comp,sorting=none,doi=false,isbn=false]{biblatex}
\AtEveryBibitem{\clearfield{pages}} \DeclareFieldFormat[article,misc,book]{title}{\mkbibbold{#1}}
\renewbibmacro{in:}{}
\hypersetup{colorlinks=true,linkcolor=blue,citecolor=blue,urlcolor=blue}
\usepackage[a4paper, total={6.5in, 9.5in}]{geometry}
\newcommand\numberthis{\addtocounter{equation}{1}\tag{\theequation}}

\renewcommand{\thesection}{\arabic{section}}
\titleformat{\section}[block]
  {\titlerule\addvspace{4pt}\normalfont\fontsize{14}{16}\bfseries}
  {\thesection.\enspace}{0pt}{\large #1}[\vspace{2pt}\titlerule]
\definecolor{airforceblue}{rgb}{0.36, 0.48, 0.84}
\definecolor{purplemath}{rgb}{0.5, 0, 0.5}

\makeatletter
\def\@xfootnote[#1]{%
  \protected@xdef\@thefnmark{#1}%
  \@footnotemark\@footnotetext}
\makeatother

\begin{document}

\begin{titlepage}

\setcounter{page}{1} \baselineskip=15.5pt \thispagestyle{empty}

\hfill DESY-23-143

\bigskip\
\begin{center}
{\fontsize{16}{30}\selectfont \bf Induced gravitational waves and baryon asymmetry\\\vspace{10pt}
fluctuations from primordial black hole formation}
\end{center}

\begin{NoHyper}
\vspace{0.5cm}
\begin{center}
{\fontsize{14}{30}\selectfont
Chiara Altavista$^{\star}$\footnote[]{$^\star$ altavista.1844243@studenti.uniroma1.it}
and Juli\'an Rey$^\dagger$ \footnote[]{$^\dagger$ julian.rey@desy.de}}
\end{center}
\end{NoHyper}

\begin{center}
\vskip 8pt
$^\star$\textsl{Dipartimento di Fisica, Sapienza Università di Roma, Piazzale Aldo Moro 5, 00185, Roma, Italy}\\
$^\dagger$\textsl{Deutsches Elektronen-Synchrotron DESY, Notkestr. 85, 22607 Hamburg, Germany}

\end{center}

\vspace{0.6cm}
\hrule
\vspace{0.4cm}
\noindent {\large \textbf{Abstract}} \\[0.3cm]
\begin{onehalfspacing}
\noindent We consider black hole formation due to the gravitational collapse produced by large density fluctuations during an epoch of reheating with a stiff equation of state and calculate the induced gravitational wave spectrum. By considering the existing bounds on the total energy density of gravitational waves today, we find constraints on the parameter space of this scenario. We also calculate the lepton asymmetry generated by metric perturbations via the chiral gravitational anomaly present in the Standard Model and find that, once the electroweak sphaleron processes have taken place, the large spectrum of scalar perturbations responsible for black hole formation induces a peak in the baryon asymmetry fluctuations on small scales.\par
\end{onehalfspacing}
\vspace{0.4cm}
\hrule

\vspace{0.1in}
\tableofcontents
\vspace{0.2in}
\titlerule

\end{titlepage}

\phantomsection
\section*{Introduction}
\addcontentsline{toc}{section}{Introduction}

The baryon-to-photon ratio observed in CMB experiments \cite{Planck:2018vyg} as well as in measurements of the abundance of light elements produced during Big Bang nucleosynthesis (see e.g.\,\cite{Mossa:2020gjc}), leads to the requirement of some mechanism, known as baryogenesis, to produce an asymmetry between matter and antimatter in the early Universe. One possibility that has been extensively considered is that of leptogenesis \cite{Davidson:2008bu}, which consists in generating a lepton asymmetry at early stages which is later converted to a baryon asymmetry via sphaleron transitions, non-perturbative processes that take place in the electroweak sector of the Standard Model. Typical models require extending the particle content of the theory by adding, for instance, right-handed neutrinos \cite{Fukugita:1986hr}. An alternative possibility is gravitational leptogenesis, which consists in exploiting the chiral gravitational anomaly \cite{Kimura:1969iwz,Alvarez-Gaume:1983ihn} present in the Standard Model (that is, in the presence of only left-handed neutrinos\footnote{Even if right-handed neutrinos are added to the Standard Model, this anomaly is still present once they are integrated out, at sufficiently low energies \cite{Alexander:2004us}.}),
\begin{equation}
\label{eq:current}
\nabla_\mu J^\mu_{\rm L}=-\frac{N_{{\rm L}-{\rm R}}}{24(4\pi)^2}R\tilde{R},
\end{equation}
where $R\tilde{R}=(1/2)\tensor{\epsilon}{^{\mu\nu}_{\rho\sigma}}R_{\mu\nu\alpha\beta}R^{\rho\sigma\alpha\beta}$ denotes the contraction between the Riemann tensor and its dual, and the factor $N_{{\rm L}-{\rm R}}=3$ arises from the difference between the number of left- and right-handed neutrinos in the theory.\footnote{In general, all chiral fermions in the theory contribute to $N_{{\rm L}-{\rm R}}$, with the contribution of each particle weighted by its corresponding ${\rm B}-{\rm L}$ factor (see e.g.\,\cite{Adshead:2017znw}), but since the Standard Model contains an equal amount of left- and right-handed quarks, their contribution vanishes and B drops out. The same argument applies to charged leptons.} By expanding the right-hand side of the above equation to quadratic order in perturbations, one obtains a term proportional to the product $h_{ij}\phi$, as well as terms quadratic in $h_{ij}$, where $\phi$ denotes the Newtonian potential and $h_{ij}$ the transverse, traceless tensor perturbation of the metric. From the structure of the resulting terms one can check that, in order for the mechanism to work, a chiral gravitational wave background is required, as one would expect from the Sakharov conditions \cite{Sakharov:1967dj}. Such a spectrum can be generated, for instance, during inflation, in models in which the inflaton $\varphi$ contains a CP-odd component and couples to gravity through a term of the form $f(\varphi)R\tilde{R}$ \cite{Alexander:2004us}.\footnote{See \cite{Kamada:2019ewe} and the references therein for a discussion about the caveats of this mechanism, in particular the presence of ghost modes, and some proposed solutions.}

It was recently suggested in \cite{Maroto:2022xrv} that, due to the fact that inflation is a stochastic process, there is actually no need to invoke the presence of these couplings to produce a lepton asymmetry via the above mechanism. Indeed, due to the dependence of the anomaly on the stochastic variables $\phi$ and $h_{ij}$, we expect a non-vanishing variance for the lepton number density $\langle |n_{\rm L}|^2\rangle$ to be generated in different patches as inflation progresses. The average of the root-mean-square variance over some particular region therefore quantifies the expected asymmetry. Since these fluctuations are generated during inflation (out of thermal equilibrium) and the anomaly violates C, CP, and L (with the electroweak sphalerons providing the required B violation later on), the Sakharov conditions are fulfilled. As shown in \cite{Maroto:2022xrv}, however, when this average is computed over the entire observable Universe today, the resulting asymmetry turns out to be extremely suppressed, rendering the proposal incapable of producing baryogenesis. Although unable to explain the observed background value of the baryon asymmetry, the presence of the chiral gravitational anomaly would nevertheless lead to unavoidable fluctuations in the baryon number density on sufficiently large scales,\footnote{As explained in \cite{Maroto:2022xrv}, the baryon asymmetry survives only on large scales due to the matter-antimatter annihilation processes taking place on small patches. We discuss this in detail in Section \ref{sec:smooth_lepton}.} which could potentially be used as an observable to probe different models of inflation. We anticipate, however, that in the particular scenario discussed here the fluctuations are much smaller than the observed background value, so measuring them would likely require some additional enhancement mechanism.

The calculation presented in \cite{Maroto:2022xrv} makes use of the $h_{ij}\phi$ term that arises after expanding eq.\,(\ref{eq:current}) in perturbations, so that the size of the fluctuations in the baryon asymmetry is proportional to the amplitude of both the tensor and scalar power spectrum, and becomes negligible if the tensor-to-scalar ratio $r$ is sufficiently small. A natural question is therefore whether an enhancement in either of these quantities could significantly increase the size of these fluctuations. Such an enhancement can be obtained, for instance, in single-field models of inflation that aim to generate a significant population of primordial black holes by introducing an inflection point in the potential (see e.g.\,\cite{Garcia-Bellido:2017mdw,Ballesteros:2017fsr,Ballesteros:2020qam} for particular implementations of this mechanism). In this class of models, the presence of the inflection point leads to a phase of ultra-slow-roll that enhances the curvature power spectrum, producing large energy overdensities once the perturbations re-enter the horizon after the end of inflation, which in turn induce gravitational collapse, generating the black holes. These black holes could account for the entirety of the observed dark matter provided their masses lie in the range
\begin{equation}
\label{eq:mass_range}
10^{-16}M_\odot\lesssim M_{\rm PBH}\lesssim 10^{-11}M_\odot,
\end{equation}
where the upper bound comes from microlensing observations \cite{Niikura:2017zjd} and the lower one from their Hawking evaporation \cite{Carr:2016hva,Arbey:2019vqx,Laha:2020ivk,Berteaud:2022tws}. Such a large scalar power spectrum would also source gravitational waves at second order in perturbations after the end of inflation \cite{Tomita:1967lol,Ananda:2006af,Baumann:2007zm}, leading to a peaked gravitational wave signal which would also contribute to the variance of the lepton number density and which is completely independent of the primordial tensor spectrum arising from inflation. The spectrum of gravitational waves induced by large scalar perturbations during a radiation era has been extensively studied in the literature \cite{Ananda:2006af,Baumann:2007zm,Espinosa:2018eve,Kohri:2018awv}. However, since the resulting signal depends on the evolution of the scalar perturbations after they re-enter the horizon, we expect the result to change if the background fluid has a different equation of state. The case in which the Universe is dominated by non-relativistic matter before transitioning to the radiation era was studied in \cite{Kohri:2018awv,Inomata:2019ivs,Inomata:2019zqy}, and the scenario with a general background was considered in \cite{Domenech:2019quo}, where an enhancement of the spectrum for background fluids with stiff equations of state (that is, for $w\lesssim 1$, with $p=w\rho$) was reported. A similar setup with a scalar field rolling down an exponential potential was studied in \cite{Domenech:2020scalar}.

This paper has two objectives. The first is to connect the enhancement of the induced gravitational wave spectrum for stiff background equations of state to the abundance of primordial black holes.\footnote{See also \cite{Liu:2023pau,Gomez-Aguilar:2023bej} for recent work in similar directions.} We consider a scenario in which primordial black holes form during a reheating stage\footnote{Throughout the paper we refer to this early stiff epoch as reheating, since we assume it takes place immediately after inflation ends. However, we remark that our results do not depend on the origin of this epoch, which we remain agnostic about. In particular, it does not necessarily have to come from the dynamics of the inflaton.} with an equation of state parameter $1/3<w<1$ before the Universe transitions to the radiation era at some temperature $T_r$. We find that, depending on the specific values of the equation of state, the transition temperature, and the scale at which the peak in the scalar power spectrum is located (therefore, on the mass and abundance of the black holes that form), the enhancement of the signal can be  large enough to violate the existing bounds on the total gravitational wave energy density derived from CMB observations and the abundance of light elements produced during Big Bang nucleosynthesis \cite{Clarke:2020bil,Pagano:2015hma}, effectively reducing the parameter space of this scenario. We perform the calculation of the gravitational wave spectrum for two cases, one in which the transition between epochs is instantaneous, and one in which the stiff fluid gradually decays into radiation, and show that the resulting constraints on the parameter space for PBH formation depend only mildly on the smoothness of the transition. We improve upon the results of \cite{Domenech:2019quo} by implementing a matching procedure for the transfer function of the scalar perturbations and the Green's function of the tensor modes in the sudden transition case akin to the one presented in \cite{Kohri:2018awv} for the case of matter-domination, effectively taking into account the full time evolution of both quantities. We also perform a fully numerical calculation of the scalar transfer function and the tensor Green's functions in the gradual transition case, in contrast to \cite{Inomata:2019zqy}.\footnote{The difference between the numerical transfer function and the phenomenological fit of \cite{Inomata:2019zqy} is depicted in Fig.\,3 of that paper for the case of a matter-dominated era with $w=0$. We focus instead on the case $w>1/3$.} The second objective is to calculate the baryon asymmetry fluctuations induced by the chiral gravitational anomaly in eq.\,(\ref{eq:current}) for the same scenario.\footnote{The mechanism studied here is different from leptogenesis via PBH evaporation, see e.g.\,\cite{Baumann:2007yr}.} We extend the results of \cite{Maroto:2022xrv} in two significant ways. The first is that we consider a peaked scalar power spectrum responsible for PBH formation, as opposed to a scale-invariant one, thereby enhancing the asymmetry reported there. The second one is that we also consider the purely scalar contribution to the asymmetry due to induced gravitational waves, effectively removing one of the essential ingredients in \cite{Maroto:2022xrv}, namely, the need to have a non-vanishing gravitational wave background generated during inflation. We once again remark that the mechanism studied here is unable to produced the observed baryon asymmetry of the Universe, but it does allow us to predict a spectrum of fluctuations in this quantity which would be present in any model of PBH formation from single-field inflation using an inflection point in the potential, assuming only the matter content of the Standard Model. Moreover, the machinery presented here could be used to compute these fluctuations in other models of gravitational leptogenesis, such as the one in \cite{Alexander:2004us}.

The paper is structured as follows. In Section \ref{sec:lepton_anomaly} we expand eq.\,(\ref{eq:current}) in perturbations and derive the corresponding expressions for the lepton number density at each order. In Section \ref{sec:bh_masses} we derive the expressions for the black hole mass and abundance assuming that they form during a reheating stage with a stiff equation of state, and discuss the relevant constraints on the parameter space. In Section \ref{sec:induced_waves} we derive the induced gravitational wave spectrum when the transition between the reheating stage and the radiation era is instantaneous, and determine the constraint on the PBH masses due to the aforementioned bounds on the gravitational wave energy density. In Section \ref{sec:gradual} we repeat the calculation by assuming that the stiff fluid gradually decays into radiation. Finally, in Section \ref{sec:smooth_lepton} we determine the baryon asymmetry produced by the chiral gravitational anomaly in this scenario.

\section{Gravitational lepton anomaly}
\label{sec:lepton_anomaly}

In this section we expand eq.\,(\ref{eq:current}) to third order in perturbations and find expressions for the lepton number density $n_{\rm L}$ at each order. Throughout the paper we denote second-order perturbations of the metric with bold symbols. The perturbed FLRW metric is, in conformal time $d\eta=dt/a$,
\begin{equation}
\label{eq:metric}
ds^2=a^2\Big\{-(1+2\phi+\Phi)d\eta^2+\Big[(1-2\phi-\Psi)\delta_{ij}+\frac{1}{2}(\partial_i{\bm E}_j+\partial_j{\bm E}_i)+h_{ij}+\frac{1}{2}{\bm h}_{ij}\Big]dx^idx^j\Big\},
\end{equation}
where we have fixed the Newtonian gauge, so that $E=B={\bm E}={\bm B}=0$, we have assumed that the first-order vector perturbations vanish, $E_i=B_i=0$, and we have used a second-order vector gauge-transformation to set ${\bm B}_i=0$. The second-order vector perturbation ${\bm E}_i$ cannot be set to zero, since it is sourced by terms quadratic in first-order scalar perturbations by virtue of Einstein's equations. We have additionally assumed that no anisotropic stress is present, so that the two first-order Newtonian potentials are equal to each other, $\phi=\psi$. The difference between the two Newtonian potentials at second order $\Phi-\Psi$ does not vanish in the absence of anisotropic stress, since it is sourced by terms quadratic in first-order scalars.

The quantity $R\tilde{R}$ which appears on the right-hand side of eq.\,(\ref{eq:current}) can be easily checked to vanish at the background level. This quantity can be calculated at each order in perturbations by direct expansion using the above metric. However, the result can also be obtained by noting that every term in the expansion must contain four derivatives (with two coming from each factor of $R$) and the indices can only be saturated in a limited number of ways. For instance, this quantity vanishes at leading order\footnote{We use the terminology LO, NLO, and NNLO for the expansion in powers of the metric perturbations. Note, however, that this does not imply anything about the relative amplitude of each term. As we will see, the dominant term changes depending on the scale considered.} (LO), since the only perturbations available are $\phi$ and $h_{ij}$, and every possible contraction (for instance, $\epsilon_{ijk}\partial_i\partial_j\partial_k\phi$) vanishes due to the antisymmetry of $\epsilon_{ijk}$. At the next-to-leading order (NLO) we have only one possible term mixing scalar and tensor perturbations,
\begin{equation}
\label{eq:scalar-tensor-first}
R\tilde{R}\supset \epsilon_{ijk}\partial_\ell\partial_i\phi\partial_j h'_{k\ell},
\end{equation}
where primes denote derivatives with respect to conformal time ($\prime=d/d\eta$), as well as several possible terms mixing two tensor perturbations, such as
\begin{equation}
\label{eq:tensor_squared}
R\tilde{R}\supset \epsilon_{ijk} \partial_m h'_{i\ell}\partial_m\partial_k h_{j\ell}.
\end{equation}
We generically expect these terms to be suppressed with respect to the scalar-tensor ones (since the stochastic gravitational wave background produced during inflation is much smaller than the scalar power spectrum) and thus we do not consider them.

At NNLO, we focus on the situation in which the first-order tensor perturbation $h_{ij}$ is negligible and the only relevant contribution comes from the scalar-induced ${\bm h}_{ij}$, which is the case in models of PBH formation from single-field inflation. In this case, all terms containing only scalar modes (such as $\epsilon_{ijk}\partial_i\phi\partial_j\phi\partial_k\phi'$ or $\epsilon_{ijk}\partial_i\partial_j\phi\partial_k\Phi'$) can be easily seen to vanish due to the antisymmetry of $\epsilon_{ijk}$. Similarly, the scalar-vector terms vanish due to the fact that terms such as $\epsilon_{ijk}\partial_i\phi'\partial_j{\bm E}'_k$ are always accompanied by $\epsilon_{ijk}\partial_i\phi'\partial_k{\bm E}'_j$, since only the symmetric combination $\partial_i{\bm E}_j+\partial_j{\bm E}_i$ appears in the metric. Thus, the only relevant term at this order is the scalar-tensor one
\begin{equation}
\label{eq:scalar-tensor-second}
R\tilde{R}\supset \epsilon_{ijk}\partial_\ell\partial_i\phi\partial_j {\bm h}'_{k\ell},
\end{equation}
since terms quadratic in ${\bm h}_{ij}$ are of higher order, and we neglect terms mixing $h_{ij}$ and ${\bm h}_{ij}$, which are subdominant with respect to the above contribution.

We conclude that, at NLO, the only relevant term is the one in eq.\,(\ref{eq:scalar-tensor-first}), assuming that the first-order gravitational wave background produced during inflation is suppressed with respect to the scalar one at the scales of interest. The numerical prefactor can be obtained by explicit calculation, and the result is
\begin{equation}
\label{eq:NLOres}
R\tilde{R}=-\frac{8}{a^4}\epsilon_{ijk}\partial_\ell\partial_i\phi\partial_j h'_{k\ell}.
\end{equation}
On the other hand, at NNLO only the term in eq.\,(\ref{eq:scalar-tensor-second}) contributes, assuming vanishing first-order tensor modes. The prefactor can be obtained by simply substituting $h_{ij}\rightarrow \frac{1}{2}{\bm h}_{ij}$ in eq.\,(\ref{eq:NLOres}),
\begin{equation}
\label{eq:NNLO_first}
R\tilde{R}=-\frac{4}{a^4}\epsilon_{ijk}\partial_\ell\partial_i\phi\partial_j {\bm h}'_{k\ell}.
\end{equation}
We remark that the NNLO terminology in no way implies that the term in eq.\,(\ref{eq:NNLO_first}) is smaller than the one in eq.\,(\ref{eq:NLOres}). The reason is that, since ${\bm h}_{ij}$ is sourced by scalar perturbations, its amplitude will depend on their initial conditions and time evolution, which can be very different from those of the unsourced $h_{ij}$.

The left-hand side of eq.\,(\ref{eq:current}) can be expanded by writing $J^\mu=(a^{-1}n_{\rm L},{\bm 0})$ and using the following identity,
\begin{equation}
\nabla_\mu J^\mu=\frac{1}{\sqrt{-g}}\partial_\mu(\sqrt{-g}J^\mu).
\end{equation}
We therefore obtain, for the NLO term,
\begin{equation}
\label{eq:nlo_ori}
a^3n_{\rm L}=\frac{1}{16\pi^2}\int \epsilon_{ijk}\partial_j\partial_\ell\phi\partial_kh^\prime_{i\ell} d\eta.
\end{equation}
We can expand the perturbations in Fourier modes
\begin{align}
h_{ij}({\bm x})&=\int\frac{d^3 {\bm k}}{(2\pi)^3}e^{i {\bm k}\cdot {\bm x}}(h_k^+e_{ij}^++h_k^\times e_{ij}^\times),\\
\phi({\bm x})&=\int\frac{d^3 {\bm k}}{(2\pi)^3}e^{i {\bm k}\cdot {\bm x}}\phi_k,
\end{align}
where $e_{ij}^s$ denotes the two transverse, traceless polarization tensors, defined via
\begin{align}
\label{eq:pol_tensors}
e^+_{ij}=\frac{1}{\sqrt{2}}(v_iv_j-\bar{v}_i\bar{v}_j),\quad		
e^\times_{ij}=\frac{1}{\sqrt{2}}(v_i\bar{v}_j+\bar{v}_iv_j),
\end{align}
where $\bm{v}$ and $\bar{\bm{v}}$ are two unit vectors satisfying $\bm{k}\cdot\bm{v}=\bm{k}\cdot\bar{\bm{v}}=\bm{v}\cdot\bar{\bm{v}}=0$. We can therefore regard $e_{ij}^s$ as a ${\bm k}$-dependent quantity. We then find
\begin{equation}
\label{eq:nlo_fourier}
a^3n_{\rm L}=\frac{1}{16\pi^2}\epsilon_{ijk}\sum_s\int d\eta\int\frac{d^3 {\bm p}}{(2\pi)^3}\int\frac{d^3 {\bm q}}{(2\pi)^3}e^{i({\bm p}+{\bm q})\cdot {\bm x}}(iq_j)(iq_\ell)\phi_q(ip_k)h_p^{s\prime }e_{i\ell}^s({\bm p}).
\end{equation}
The mean value of this quantity clearly vanishes, as can be easily checked by expanding $\phi_k$ and $h_k^s$ in terms of creation and annihilation operators. The variance, however, does not, and will be computed in Section \ref{sec:smooth_lepton}.

To find the expression for the NNLO term in Fourier space it is necessary to solve the equation of motion for the second-order tensor modes induced by scalar fluctuations, given, in momentum space and in the absence of anisotropic stress, by (see e.g.\,\cite{Baumann:2007zm,Kohri:2018awv,Espinosa:2018eve,Inomata:2019ivs})
\begin{equation}
\label{eq:tensor_eom}
{\bm h}_k^{s\prime\prime}+2\mathcal{H}{\bm h}_k^{s\prime}+k^2 {\bm h}_k^s=S^s_k,
\end{equation}
where $\mathcal{H}=a'/a$ denotes the conformal Hubble factor and the source term is, in the Newtonian gauge,
\begin{equation}
S_k^s=\int\frac{d^3p}{(2\pi)^3}\Big[{\bm p}\cdot {\bm e}^s({\bm k})\cdot {\bm p}\Big]\bigg[8\phi_p\phi_{|{\bm k}-{\bm p}|}+\frac{16}{3(1+p/\rho)}\bigg(\phi_p+\frac{\phi'_p}{\mathcal{H}}\bigg)\bigg(\phi_{|{\bm k}-{\bm p}|}+\frac{\phi'_{|{\bm k}-{\bm p}|}}{\mathcal{H}}\bigg)\bigg],
\end{equation}
where $p$ and $\rho$ are the background pressure and energy density, and we have used the boldface ${\bm e}^s({\bm k})$ to denote the matrix $e_{ij}^s({\bm k})$. Explicitly, we have ${\bm p}\cdot {\bm e}^s({\bm k})\cdot {\bm p}\equiv p_i e_{ij}^s({\bm k}) p_j$. The solution to eq.\,(\ref{eq:tensor_eom}) is given by
\begin{equation}
{\bm h}_k^s(\eta)=\int_0^\eta G_k(\eta,\eta')S_k^s(\eta')d\eta',
\end{equation}
where $G_k(\eta,\eta')$ is the Green's function of the differential equation, and we have assumed that inflation ends at $\eta=0$ and the gravitational waves induced up to that point are negligible. This solution can be rewritten as
\begin{equation}
\label{eq:h_express}
{\bm h}_k^s(\eta)=\frac{1}{k^2}\int\frac{d^3{\bm p}}{(2\pi)^3}\Big[{\bm p}\cdot {\bm e}^s({\bm k})\cdot {\bm p}\Big]\mathcal{R}_p\mathcal{R}_{|{\bm k}-{\bm p}|}I_k(\eta,p,|{\bm k}-{\bm p}|),
\end{equation}
where
\begin{equation}
\label{eq:i_function}
I_k(\eta,p,|{\bm k}-{\bm p}|)=\bigg(\frac{3+3w}{5+3w}\bigg)^2\int_0^\eta kG_k(\eta,\eta')Q(\eta',p,|{\bm k}-{\bm p}|) kd\eta',
\end{equation}
with
\begin{align}
Q(\eta,p,|{\bm k}-{\bm p}|)&=8T_\phi(p\eta)T_\phi(|{\bm k}-{\bm p}|\eta)\nonumber\\
&+\frac{16}{3(1+p/\rho)}\bigg(T_\phi(p\eta)+\frac{T_\phi'(p\eta)}{\mathcal{H}}\bigg)\bigg(T_\phi(|{\bm k}-{\bm p}|\eta)+\frac{T_\phi'(|{\bm k}-{\bm p}|\eta)}{\mathcal{H}}\bigg).
\end{align}
Here we have used the definition of the scalar transfer function $\phi_k(\eta)\equiv T_\phi(k\eta)\phi_k(0)$, obtained by solving the equation of motion for $\phi_k$, as well as the fact that the initial value of the Newtonian potential $\phi_k(0)$ is related to the frozen curvature perturbation $\mathcal{R}_k$ on superhorizon scales via \cite{Baumann:Cosmo2022}
\begin{equation}
\label{eq:newt_superhorizon}
\phi_k(0)=\frac{3+3w}{5+3w}\mathcal{R}_k,
\end{equation}
where $w$ is the equation of state of the Universe at the time at which the initial conditions are imposed (that is, shortly after the end of inflation). In Section \ref{sec:induced_waves} we will give explicit expressions for the Green's function $G_k(\eta,\eta')$ and the transfer function.

At NNLO, the lepton number density is given by
\begin{equation}
a^3n_{\rm L}=\frac{1}{32\pi^2}\int \epsilon_{ijk}\partial_j\partial_\ell\phi\partial_k{\bm h}^\prime_{i\ell}d\eta,
\end{equation}
where we reiterate that we have neglected all terms involving the first-order tensor modes $h_{ij}$. Using eq.\,(\ref{eq:h_express}) we can write this quantity as
\begin{align}
a^3n_{\rm L}&=\frac{1}{32\pi^2}\bigg(\frac{3+3w}{5+3w}\bigg)\epsilon_{ijk}\sum_s\int d\eta\int\frac{d^3{\bm p}}{(2\pi)^3}\int\frac{d^3{\bm q}}{(2\pi)^3}e^{i({\bm p}+{\bm q})\cdot {\bm x}}(iq_j)(iq_\ell)(ip_k)e_{i\ell}^s({\bm p})\nonumber\\
&\quad
T_\phi(q\eta)
\frac{1}{p^2}\int\frac{d^3{\bm k}}{(2\pi)^3}\bigg[{\bm k}\cdot {\bm e}^s({\bm p})\cdot {\bm k}\bigg]I_p'(\eta,k,|{\bm p}-{\bm k}|) \mathcal{R}_{\bm q}\mathcal{R}_{\bm k}\mathcal{R}_{{\bm p}-{\bm k}}.
\label{eq:nnlo_ori}
\end{align}
As we will see in Section \ref{sec:smooth_lepton}, the mean value of this quantity once again vanishes, but the variance does not. In general, $n_{\rm L}$ is given by the sum of both the NLO and NNLO contributions.

\section{Black hole masses and abundance}
\label{sec:bh_masses}

Black holes form when sufficiently large perturbations re-enter the horizon after the end of inflation. Their abundance can be described using the Press-Schechter formalism, which states that collapse occurs only when the density contrast $\delta=\delta\rho/\rho$ is above a critical value $\delta_c$,
\begin{align}
\label{eq:press}
\beta=\frac{1}{\sqrt{2\pi\sigma^2}}\int_{\delta_c}^\infty e^{-\delta^2/(2\sigma^2)}d\delta,
\end{align}
where $\beta=\rho_{\rm PBH}/(\gamma\rho)$ is the ratio of the energy density in a Hubble patch that ends up in the form of PBHs to the total energy density at the time of collapse, with $\gamma$ an $\mathcal{O}(1)$ factor encoding the efficiency of the collapse.\footnote{We set $\gamma= 0.2$ throughout this work \cite{Carr:1975qj}, but we remark that this number could depend on the equation of state parameter $w$ at the time at which the black holes form.} In the above expression we have assumed the probability distribution of the fluctuations to be Gaussian and the smoothed variance of the density contrast $\sigma$ can be related to the dimensionless primordial spectrum of curvature perturbations $\mathcal{P}_\mathcal{R}$ by means of the gradient expansion \cite{Harada:2015yda}
\begin{equation}
\label{eq:variance}
\sigma^2=\frac{4(1+w)^2}{(5+3w)^2}\int\frac{dq}{q}\bigg(\frac{q}{k}\bigg)^4\mathcal{P}_\mathcal{R}(q)W^2(q/k),
\end{equation}
where $W(x)=e^{-x^2/2}\sqrt{2/\pi}$ is a window function, which we take as a Gaussian. The dependence of the critical value $\delta_c$ in eq.\,(\ref{eq:press}) on the equation of state parameter $w$ was analytically estimated in \cite{Harada:2013epa} to be
\begin{equation}
\label{eq:delta_c}
\delta_c=\frac{3(1+w)}{5+3w}\sin^2\bigg(\frac{\pi\sqrt{w}}{1+3w}\bigg).
\end{equation}
Note that this formula is not valid for $w\ll1/3$, since in this case the non-sphericity and angular momentum of the collapsing cloud, neglected in \cite{Harada:2013epa}, become relevant, see \cite{Harada:2016mhb,Harada:2017fjm}.

Recent studies have focused on using the peak theory formalism instead of Press-Schechter to calculate the PBH abundance, see e.g.\,\cite{Riccardi:2021rlf}. Similarly, the fact that in models of PBH formation the probability distribution deviates significantly from the Gaussian approximation is well-understood \cite{Taoso:2021uvl}. However, as noted in \cite{Taoso:2021uvl,Riccardi:2021rlf}, the change in abundance induced by taking into account these modifications can always be compensated by multiplying the power spectrum by an $\mathcal{O}(1)$ factor, since eq.\,(\ref{eq:press}) depends exponentially on $\mathcal{P}_\mathcal{R}$. Since the induced gravitational wave spectrum depends quadratically on $\mathcal{P}_\mathcal{R}$, these corrections will not heavily impact the results presented in this work, and in any case can be incorporated by a simple rescaling of the power spectrum. The same argument applies to the threshold $\delta_c$, which strictly speaking should be calculated by finding the local maxima of the compaction function, a quantity that measures how close the overdensity is to fulfilling the hoop conjecture, see e.g.\,\cite{Germani:2023ojx}. The specific value of $\delta_c$ will not be particularly important for our arguments, since the only issue relevant for us is the value of $\mathcal{P}_\mathcal{R}$ necessary to obtain $\Omega_{\rm PBH}^0=\Omega_{\rm DM}^0$ (that is, for PBHs to form all of the dark matter today), which is always of order $\mathcal{P}_\mathcal{R}\sim 10^{-2}$. We take eqs.\,(\ref{eq:press}, \ref{eq:delta_c}) simply as a convenient benchmark, and we remark that the main results of this paper do not rely on the use of the Press-Schechter formalism at all.

\begin{figure}[t]
\begin{center}
$$\includegraphics[width=.48\textwidth]{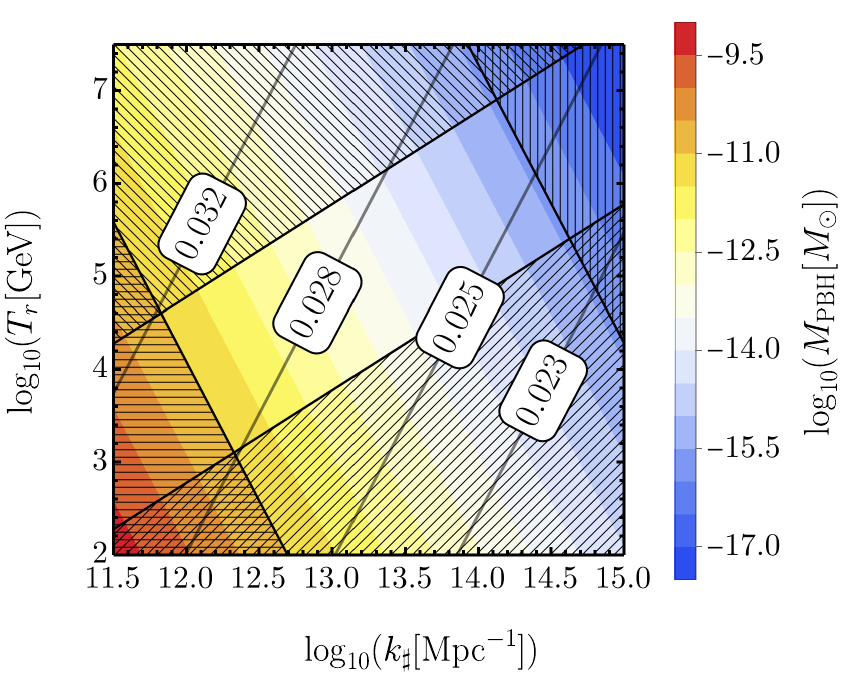}\qquad
\includegraphics[width=.48\textwidth]{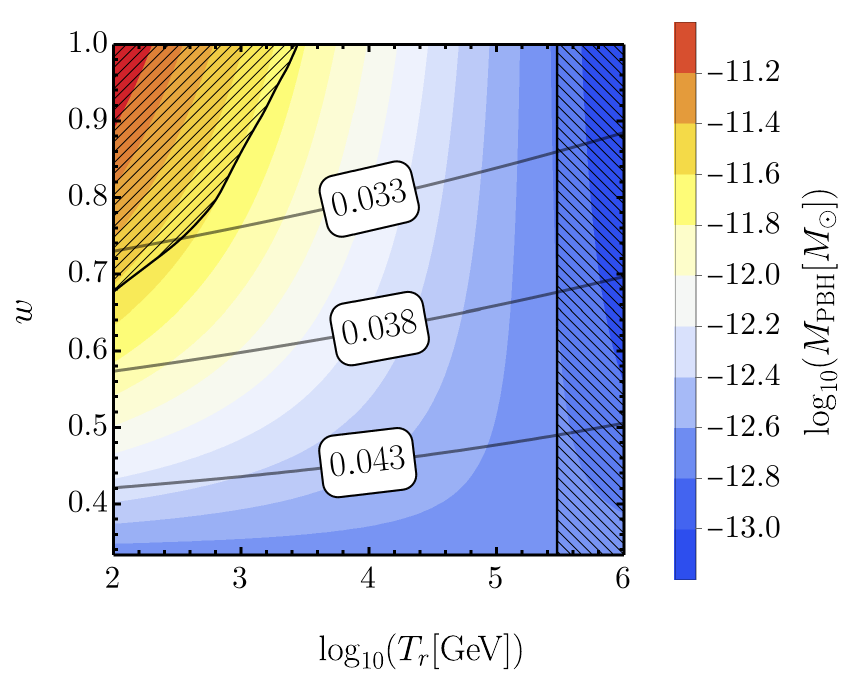}$$
\caption{\em \label{fig:abu_mass} {Left panel: contours depicting the black hole masses as a function of the reheating temperature $T_r$ and the scale $k_\sharp$ at which the peak in the power spectrum is located for $w=1$, together with the amplitude of the power spectrum $\mathcal{A}_\sharp$ necessary to obtain $f_{\rm PBH}=1$ (labeled solid lines). The horizontal and vertical shaded regions represent the constraints on the PBH masses in eq.\,(\ref{eq:mass_range}). The region with bottom-right to top-left shading represents the re-entry constraint in eq.\,(\ref{eq:entry_con}), and the region with bottom-left to top-right shading represents the GW constraint in eq.\,(\ref{eq:gw_con}). Right panel: same as left panel, but leaving $w$ as a free parameter, for $k_\sharp=5\times 10^{12}\,{\rm Mpc}^{-1}$.
}}
\end{center}
\end{figure}

In what follows we consider a situation in which the Universe goes through a long phase of reheating after inflation with an arbitrary equation of state $w>1/3$ during which PBHs form, and then transitions instantaneously into a radiation-dominated era at temperature $T_r$. By following a procedure analogous to the one presented in \cite{Ballesteros:2019hus} for the case $w=0$, we can write the present fraction of dark matter in the form of PBHs $f_{\rm PBH}\equiv\Omega^0_{\rm PBH}/\Omega^0_{\rm DM}$ as
\begin{equation}
\label{eq:pbh_abu}
f_{\rm PBH}=\gamma\beta\frac{\Omega_{\gamma}^0}{\Omega_{\rm DM}^0}\bigg[\bigg(\frac{g_{\star s}(T_r)}{g_{\star s}(T_0)}\bigg)^{1/3}\frac{T_r}{T_0}\bigg]^{\frac{1+9w}{1+3w}}\bigg(\frac{M_p^2k^2}{T_r^4}\frac{90}{\pi^2g_\star(T_r)}\bigg)^{\frac{3w}{1+3w}},
\end{equation}
where $T_0$ is the temperature of the Universe today, $\Omega_\gamma$ the energy density in radiation, $g_\star$ denotes the effective number of relativistic degrees of freedom, and $g_{\star s}$ the effective number of degrees of freedom in entropy.\footnote{We take $g_\star(T_r)=g_{\star s}(T_r)=106.75$, $g_{\star}(T_0)=3.36$, and $g_{\star s}(T_0)=3.94$ throughout the paper \cite{Baumann:Cosmo2022}.} We have normalized the scale factor as $a_0=1$. In deriving the above equation we have assumed that collapse occurs immediately after perturbations with wavenumber $k$ re-enter the horizon, at a time $t_k$ defined by $k=a(t_k)H(t_k)$. We assume entropy is conserved only between the transition time $t_r$ and today, but not necessarily between the time of collapse and $t_r$. The mass of the PBHs is proportional to the mass contained in a Hubble patch at the time of formation, $M_{\rm PBH}=4\pi\gamma M_p^2/H$. By following a similar procedure to the one we used for $f_{\rm PBH}$, we can write
\begin{equation}
\label{eq:pbh_mass}
M_{\rm PBH}=4\pi\gamma M_p^2\bigg(\frac{\pi^2}{90}g_\star(T_r)\frac{T_r^4}{M_p^2}\bigg)^{\frac{1}{1+3w}}\bigg(\frac{1}{k^3}\frac{g_{\star s}(T_0)T_0^3}{g_{\star s}(T_r)T_r^3}\bigg)^{\frac{1+w}{1+3w}}.
\end{equation}
As a cross-check, note that, for $w=1/3$, eqs.\,(\ref{eq:pbh_abu}) and (\ref{eq:pbh_mass}) become independent of $T_r$ and, in particular, we recover the well-known scaling relation $M_{\rm PBH}\propto k^{-2}$ \cite{Ballesteros:2017fsr}.

It is clear from eqs.\,(\ref{eq:press}) and (\ref{eq:variance}) that to produce a large PBH population one requires a spectrum of curvature perturbations $\mathcal{P}_\mathcal{R}$ which is peaked at a particular scale $k$ so as not to overproduce black holes with masses outside of the unconstrained range in eq.\,(\ref{eq:mass_range}). Such a power spectrum can be obtained from single-field models of inflation in which the potential has an approximate inflection point leading to a short phase of ultra-slow-roll, see e.g.\,\cite{Ballesteros:2017fsr,Ballesteros:2020qam}. In what follows we will consider two possibilities, a flat scale-invariant spectrum with an amplitude $\mathcal{A}_\flat$ that matches the value measured on CMB scales of $\mathcal{O}(10^{-9})$,
\begin{equation}
\label{eq:pr_flat}
\mathcal{P}_\mathcal{R}(k)=\mathcal{A}_\flat,
\end{equation}
and a sharp Dirac delta spectrum with an amplitude $\mathcal{A}_\sharp$ of $\mathcal{O}(10^{-2})$ in order to produce an $\mathcal{O}(1)$ fraction of PBHs as dark matter,
\begin{equation}
\label{eq:pr_sharp}
\mathcal{P}_\mathcal{R}(k)=k_\sharp\mathcal{A}_\sharp\delta(k-k_\sharp),
\end{equation}
where we have assumed that the peak occurs at a scale $k_\sharp$. This form of the power spectrum will allow us to derive several results analytically later on. Putting everything together and considering the sharp power spectrum, we find the following expression for $\beta$ at the peak of the distribution, which is located at $k=k_\sharp/\sqrt{2}$ (as can be checked by setting $d\sigma/dk=0$),
\begin{equation}
\label{eq:beta}
\beta=\frac{1}{2}{\rm erfc}\bigg[\frac{3e}{8}\sin^2\bigg(\frac{\pi\sqrt{w}}{1+3w}\bigg)\sqrt{\frac{\pi}{\mathcal{A}_\sharp}}\bigg].
\end{equation}
Note that the total PBH abundance is obtained by integrating $f_{\rm PBH}$ over $k$, but since we are considering essentially monochromatic mass distributions, we can approximate the abundance by the value of $f_{\rm PBH}$ at its peak, $f_{\rm PBH}(k=k_\sharp/\sqrt{2})$.

The black hole mass and abundance are plotted in Fig.\,\ref{fig:abu_mass}. As anticipated, the power spectrum required to obtain $f_{\rm PBH}=1$ is of order $\mathcal{P}_\mathcal{R}\sim 10^{-2}$ over the entire parameter space, so the $w$-dependent threshold $\delta_c$ does not have a particularly strong effect. In the Figures we also show the GW constraint derived in the next Section, in eq.\,(\ref{eq:gw_con}), for a fixed value $\mathcal{A}_\sharp=0.05$. Choosing the value of $\mathcal{A}_\sharp$ necessary to obtain $f_{\rm PBH}=1$ at each point in parameter space would only marginally change this constraint. We have also included the constraint imposed by the requirement that collapse occurs in the reheating stage. That is, the mode with comoving wavenumber $k_\sharp$ that induces collapse must re-enter the horizon at most at the time of the transition,
\begin{equation}
\label{eq:entry_con}
k_\sharp>a_rH_r=\frac{T_0}{T_r}\bigg(\frac{g_{\star s}(T_0)}{g_{\star s}(T_r)}\bigg)^{1/3}\bigg(\frac{\pi^2g_\star(T_r)T_r^4}{90M_p^2}\bigg)^{1/2}.
\end{equation}
We remark that this constraint does not impede the formation of black holes, but if it is not satisfied then collapse occurs during the radiation era, and one should use the expressions for the mass and abundance for $w=1/3$ instead.

We conclude that a population of PBHs with unconstrained masses which is large enough to explain the observed dark matter abundance can only form during an early stiff era (with $w=1$) in a narrow region of parameter space, for $10^3\,{\rm GeV}<T_r<10^7\,{\rm GeV}$, depending on the location of the peak in the power spectrum. We therefore find that the new GW constraint derived in this work (see the following Section) significantly restricts the parameter space for this mechanism.

\section{Induced gravitational waves}
\label{sec:induced_waves}

In this Section we solve the equation of motion (\ref{eq:tensor_eom}) for the tensor modes induced by scalar perturbations at second order. The solution, shown in eq.\,(\ref{eq:h_express}) can be written in terms of the Green's function $G_k(\eta,\eta')$ of the differential equation and the scalar transfer function $T_\phi(k\eta)$, which we now determine. 

The Green's function can be obtained by solving the homogeneous equation
\begin{equation}
\label{eq:tensor_homo}
{\bm h}_k^{s\prime\prime}+2\mathcal{H}{\bm h}_k^{s\prime}+k^2 {\bm h}_k^s=0.
\end{equation}
Since we assume that the transition between the reheating stage with equation of state parameter $w$ and the radiation epoch occurs instantaneously at $\eta_r$, the evolution of $\mathcal{H}$ is given by
\begin{equation}
\mathcal{H} =
\begin{cases}
    \mathlarger{\frac{2}{(1+3w)\eta}}, & \text{for } \eta < \eta_r,
    \vspace{0.1in}
    \\
    \mathlarger{\frac{1}{\eta-\eta_w}}, & \text{for } \eta_r < \eta,
\end{cases}
\end{equation}
where we have chosen the integration constant $\eta_w\equiv(1-3w)\eta_r/2$ to make the function continuous at $\eta_r$. We remark that throughout the paper $w$ always denotes the equation of state parameter during the stage of reheating after inflation. We never use $w$ to denote the equation of state parameter during the radiation era, and instead write explicitly $p/\rho=1/3$ wherever necessary.

The general solution to eq.\,(\ref{eq:tensor_homo}) is
\begin{equation}
h_k^s(\eta) =
\begin{cases}
    (k\eta)^{m_\ell}\Big[A_{h\ell} J_{m_\ell}(k\eta)+B_{h\ell} Y_{m_\ell}(k\eta)\Big], & \text{for } \eta < \eta_r,
    \vspace{0.1in}
    \\
    \mathlarger{\frac{1}{k\eta-k\eta_w}}\Big[A_{hr}\sin(k\eta-k\eta_w)+B_{hr}\cos(k\eta-k\eta_w)\Big], & \text{for } \eta_r < \eta,
\end{cases}
\end{equation}
where the $\ell$ subscript denotes quantities in the reheating stage, and $r$ refers to quantities in the radiation era. We have also defined
\begin{equation}
m_\ell\equiv\frac{3(w-1)}{2(1+3w)}.
\end{equation}
The Green's function is given by
\begin{equation}
G_k(\eta,\eta')=\frac{h_1(\eta)h_2(\eta')-h_1(\eta')h_2(\eta)}{h'_1(\eta')h_2(\eta')-h_1(\eta')h'_2(\eta')},
\end{equation}
where $h_1$ and $h_2$ are any two linearly independent solutions to eq.\,(\ref{eq:tensor_homo}) and we have suppressed the $k$ and $s$ indices in $h$ for readability. Since this expression is independent of which pair of solutions is chosen, we can simply fix the integration constants in one of the regions by hand, and obtain the constants in the other epoch by imposing continuity of the solutions and their derivatives at $\eta_r$. We therefore set $A_{hr}^{1}=1$ and $B_{hr}^{1}=0$ for the first solution, together with $A_{hr}^{2}=0$ and $B_{hr}^{2}=1$ for the second one.

\begin{figure}[t]
\begin{center}
$$\includegraphics[width=.47\textwidth]{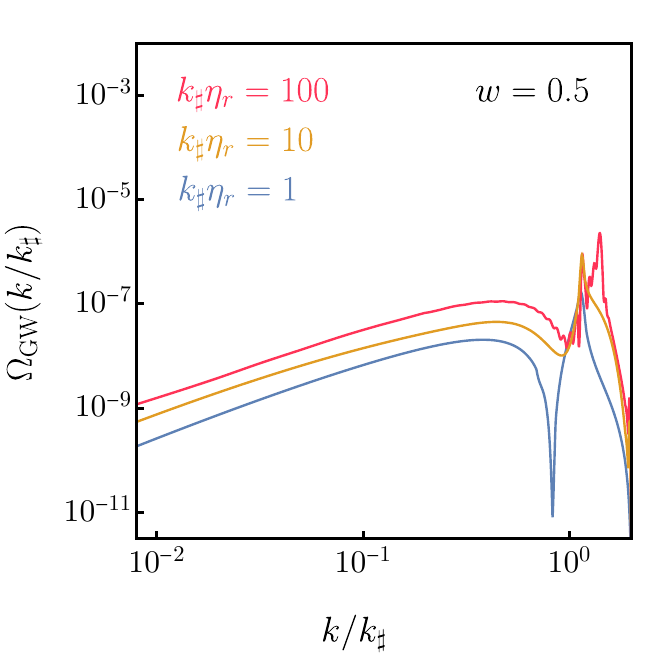}\qquad
\includegraphics[width=.47\textwidth]{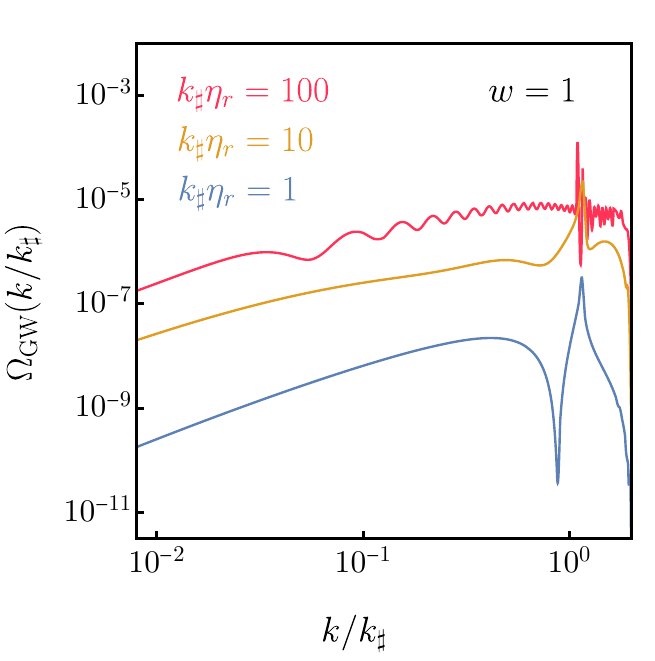}$$
\caption{\em \label{fig:gw_examples} {
Left panel: gravitational wave energy density as a function of $k/k_\sharp$ for $w=0.5$ varying the dimensionless parameter $k_\sharp\eta_r$. The energy density grows as this parameter increases. Right panel: same as left panel, for $w=1$.}}
\end{center}
\end{figure}

For the calculation of the lepton number density we need the tensor transfer function, defined by
\begin{equation}
h_k^s(\eta)= T_h(k\eta)h_k^s(0).
\end{equation}
We remind the reader that initial conditions are imposed at the end of inflation, which we choose as $\eta=0$. Assuming that tensor modes are initially frozen outside the horizon (so that we can impose the initial conditions $T_h(0)=0$ and $T_h'(0)=1$ for superhorizon modes after inflation ends), we find the following expression for the tensor transfer function
\begin{equation}
T_h(k\eta) =
\begin{cases}
    \mathlarger{\frac{\Gamma(1-m_\ell)}{2^{m_\ell}}}(k\eta)^{m_\ell}J_{-m_\ell}(k\eta), & \text{for } \eta < \eta_r,
    \vspace{0.1in}
    \\
    \mathlarger{\frac{1}{k\eta-k\eta_w}}\Big[A_{hr}\sin(k\eta-k\eta_w)+B_{hr}\cos(k\eta-k\eta_w)\Big], & \text{for } \eta_r < \eta.
\end{cases}
\end{equation}
We remark that the transfer function is unrelated to the Green's function, and in this case the constants during the reheating stage are fixed by the initial conditions mentioned above, whereas the constants $A_{hr}$ and $B_{hr}$ must be determined by imposing continuity of the solutions and their derivatives at $\eta_r$. This transfer function does not enter the calculation of the induced GW spectrum, but it is necessary for the calculations of Section \ref{sec:smooth_lepton}, so we write it here for completeness.

The equation of motion for the Newtonian potential is
\begin{equation}
\label{eq:newt_pot}
\phi_k''+3\bigg(1+\frac{p}{\rho}\bigg)\mathcal{H}\phi_k'+\frac{p}{\rho}k^2\phi_k=0,
\end{equation}
where we have assumed that the background is dominated by a perfect fluid with $\delta p=(p/\rho)\delta\rho$. We remind the reader that we are working in the Newtonian gauge and in the absence of anisotropic stress. Assuming once again that the modes are initially frozen outside the horizon, we obtain the following expression for the transfer function, valid for $w\neq 0$,
\begin{equation}
\label{eq:scalar_trans}
T_\phi(k\eta) =
\begin{cases}
   \mathlarger{\frac{\Gamma(n_\ell+1)2^{n_\ell}}{w^{n_\ell/2}}}(k\eta)^{-n_\ell}J_{n_\ell}\Big(k\eta\sqrt{w}\Big), & \text{for } \eta < \eta_r,
    \vspace{0.1in}
    \\
    \bigg(\mathlarger{\frac{\sqrt{3}}{k\eta-k\eta_w}}\bigg)^2\bigg\{A_{\phi r}\bigg[\sin\bigg(\mathlarger{\frac{k\eta-k\eta_w}{\sqrt{3}}}\bigg)+\bigg(\mathlarger{\frac{\sqrt{3}}{k\eta-k\eta_w}}\bigg)\cos\bigg(\mathlarger{\frac{k\eta-k\eta_w}{\sqrt{3}}}\bigg)\bigg]+ &\\
    \;\;\;\;\qquad\qquad\quad +B_{\phi r}\bigg[\cos\bigg(\mathlarger{\frac{k\eta-k\eta_w}{\sqrt{3}}}\bigg)-\bigg(\mathlarger{\frac{\sqrt{3}}{k\eta-k\eta_w}}\bigg)\sin\bigg(\mathlarger{\frac{k\eta-k\eta_w}{\sqrt{3}}}\bigg)\bigg]\bigg\}, &  \text{for } \eta_r < \eta,
\end{cases}
\end{equation}
where
\begin{equation}
n_\ell=\frac{5+3w}{2(1+3w)}=1-m_\ell
\end{equation}
and the constants $A_{\phi r}$ and $B_{\phi r}$ can once again be found by matching the solutions and their derivatives at $\eta_r$.

With these ingredients in hand we can calculate the $I_k$ function in eq.\,(\ref{eq:i_function}). The GW energy density is given by \cite{Maggiore:1900zz}
\begin{equation}
\label{eq:omega_gw}
\Omega_{\rm GW}(\eta,k)=\frac{1}{24}\bigg(\frac{k}{\mathcal{H}}\bigg)^2\langle\mathcal{P}_{\bm h}(\eta,k)\rangle,
\end{equation}
where the brackets $\langle\cdots\rangle$ denote a time average, which must be taken due to the stochasticity of the signal, and $\mathcal{P}_{\bm h}$ denotes the power spectrum of ${\bm h}_{ij}$. Starting from eq.\,(\ref{eq:h_express}), we find the following expression for the power spectrum (see e.g.\,\cite{Espinosa:2018eve,Kohri:2018awv})
\begin{equation}
\label{eq:tensor_spectrum}
\mathcal{P}_{\bm h}(\eta,k)=\int_0^\infty dy\int_{|1-y|}^{1+y}dz\bigg[\frac{4y^2-(1+y^2-z^2)^2}{8yz}\bigg]^2\mathcal{P}_\mathcal{R}(ky)\mathcal{P}_\mathcal{R}(kz)I_k^2(\eta,ky,kz).
\end{equation}
Since we are interested in measuring this quantity at late times, we can take $\eta\rightarrow\infty$ in eq.\,(\ref{eq:i_function}). After pulling the solutions $h_1(\eta)$ and $h_2(\eta)$ in the Green's function out of the integral and splitting the limits of integration into the two contributions $(0,\eta_r)$ and $(\eta_r,\infty)$, we obtain the following expression, valid late into the RD era,
\begin{equation}
I_k(\eta>\eta_r)= \frac{\sin(k\eta-k\eta_w)}{k\eta-k\eta_w}(J_\ell^s+J_r^s)-\frac{\cos(k\eta-k\eta_w)}{k\eta-k\eta_w}(J_\ell^c+J_r^c),
\end{equation}
where
\begin{align}
J_\ell^s&=\bigg(\frac{3+3w}{5+3w}\bigg)^2
\bigg(\frac{\pi/2}{A_{\ell h}^2B_{\ell h}^1-A_{\ell h}^1B_{\ell h}^2}\bigg)\int_0^{\eta _r}
(k\eta')^{-2m_\ell+1}h_2^\ell(\eta')Q_\ell(\eta',ky,kz)kd\eta',\\
J_\ell^c&=\bigg(\frac{3+3w}{5+3w}\bigg)^2
\bigg(\frac{\pi/2}{A_{\ell h}^2B_{\ell h}^1-A_{\ell h}^1B_{\ell h}^2}\bigg)\int_0^{\eta _r}
(k\eta')^{-2m_\ell+1}h_1^\ell(\eta')Q_\ell(\eta',ky,kz)kd\eta',
\end{align}
and
\begin{align}
J_r^s&=\bigg(\frac{3+3w}{5+3w}\bigg)^2\int_{\eta_r}^{\infty}
(k\eta'-k\eta_w)
\cos(k\eta'-k\eta_w)
Q_r(\eta',ky,kz)kd\eta',\\
J_r^c&=\bigg(\frac{3+3w}{5+3w}\bigg)^2\int_{\eta_r}^{\infty}
(k\eta'-k\eta_w)\sin(k\eta'-k\eta_w)
Q_r(\eta',ky,kz)kd\eta'.
\end{align}
Analytical expressions can be found for the last two integrals, whereas the first two must in general be performed numerically.

\begin{figure}[t]
\begin{center}
$$\includegraphics[width=.43\textwidth]{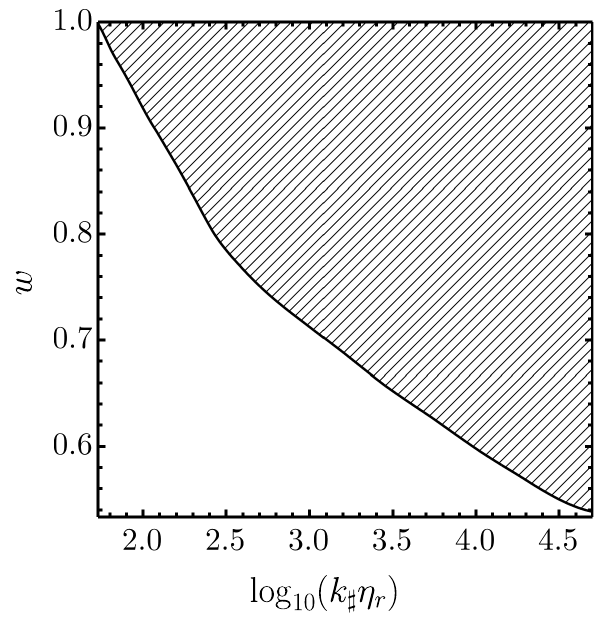}$$
\caption{\em \label{fig:gw_constraints} {
Constraint in eq.\,(\ref{eq:gw_con}) as a function of the two parameters $w$ and $k_\sharp\eta_r$. The top-right region is forbidden due to the fact that the redshift of the tensor modes during an epoch with a stiff equation of state makes the energy density grow, as illustrated in Fig.\,\ref{fig:gw_examples}.}}
\end{center}
\end{figure}

Squaring $I_k$ and taking the average as required by eq.\,(\ref{eq:omega_gw}), we find
\begin{equation}
\frac{k^2}{\mathcal{H}^2}\langle I_k^2\rangle=\frac{1}{2}\Big[(J_\ell^s+J_r^s)^2+(J_\ell^c+J_r^c)^2\Big]\equiv \frac{1}{2}J^2(y,z,k\eta_r),
\end{equation}
where we have used $\langle\sin^2(k\eta)\rangle=\langle\cos^2(k\eta)\rangle=1/2$ and $\langle\sin(k\eta)\cos(k\eta)\rangle=0$ and highlighted the dependence of $J$ on the dimensionless parameter $k\eta_r$. We can evaluate the integrals over $y$ and $z$ in eq.\,(\ref{eq:tensor_spectrum}) by using the sharp power spectrum in eq.\,(\ref{eq:pr_sharp}) necessary to obtain a significant PBH population. The result is
\begin{equation}
\label{eq:gw_energy}
\Omega_{\rm GW}(k/k_\sharp)=\mathcal{A}_\sharp^2\frac{\Omega_\gamma(T_0)}{3072}\frac{g_\star(T)}{g_\star(T_0)}\bigg(\frac{g_{\star s}(T_0)}{g_{\star s}(T)}\bigg)^{4/3}\frac{k^2}{k_\sharp^2}\bigg(4\frac{k_\sharp^2}{k^2}-1\bigg)^2J^2\bigg(\frac{k_\sharp}{k},\frac{k_\sharp}{k},k\eta_r\bigg)\Theta_{\rm H}\bigg(2\frac{k_\sharp}{k}-1\bigg),
\end{equation}
where $\Theta_{\rm H}$ denotes the Heaviside function, and we have used the fact that these gravitational waves behave as radiation at late times to evolve the result from some late time deep into the radiation era until today. This expression only depends on the two dimensionless quantities $k_\sharp/k$ and $k_\sharp\eta_r$. The transition time is, in terms of temperature,
\begin{equation}
\label{eq:eta_temp}
\eta_r=\frac{1}{T_r}\bigg(\frac{2}{1+3w}\bigg)\sqrt{\frac{90}{\pi^2g_\star(T_r)}}\bigg(\frac{g_{\star s}(T_r)}{g_{\star s}(T_0)}\bigg)^{1/3}\frac{M_p}{T_0}.
\end{equation}

In Fig.\,\ref{fig:gw_examples} we show the GW energy density from eq.\,(\ref{eq:gw_energy}) for different choices of the two parameters $w$ and $k_\sharp\eta_r$. We find that the GW abundance grows for stiff equations of state, as reported in \cite{Domenech:2019quo}, due to the fact that the tensor modes redshift more slowly than the background. We also find that the energy density grows as $k_\sharp\eta_r$ increases. The reason for this is that the largest contribution to the momentum integral in eq.\,(\ref{eq:tensor_spectrum}) comes from scales around the narrow peak in the scalar power spectrum, and therefore we should only expect the signal to be affected if the relevant modes re-enter the horizon before the transition to radiation has taken place, so that $k_\sharp>\mathcal{H}_r\propto \eta_r^{-1}$. For $k_\sharp\eta_r\ll1$, the relevant modes enter during the radiation era and the effect of the stiff epoch on the signal is washed out, as can be clearly seen in the Figure.

The total energy density can be found by integrating
\begin{equation}
\Omega_{\rm GW}=\int d\log k\;\Omega_{\rm GW}(k).
\end{equation}
As we anticipated earlier, there is a bound on this quantity arising from CMB observations and the abundance of light elements produced during Big Bang nucleosynthesis \cite{Clarke:2020bil,Pagano:2015hma},
\begin{equation}
\label{eq:gw_con}
\Omega_{\rm GW}h^2<1.8\times 10^{-6}.
\end{equation}
Since the signal grows as both $w$ and $k_\sharp\eta_r$ increase, this bound is eventually violated. In Fig.\,\ref{fig:gw_constraints} we perform a numerical scan over the parameter space, showing the forbidden region.\footnote{The small kink towards the center of the plot is a numerical artifact due to the precision of our scan and has no physical significance.} By using eq.\,(\ref{eq:eta_temp}), this bound can be written in terms of the transition temperature, and translated to a constraint on the PBH mass and abundance, shown in Fig.\,\ref{fig:abu_mass} for $w=1$ (left panel) and for $k_\sharp=5\times 10^{12}\,{\rm Mpc}^{-1}$ (right panel). These constraints are one of the main results of this paper. As anticipated at the end of the previous Section, this bound significantly limits the available parameter space for PBH formation during a stiff epoch. The constraint disappears completely for the standard radiation scenario with $w = 1/3$, where the bound is satisfied. In these Figures we have taken $\mathcal{A}_\sharp=5\times 10^{-2}$ (thereby assuming that $f_{\rm PBH}\simeq 1$) and neglected the mild dependence of this number on $w$ depicted in Fig.\,\ref{fig:abu_mass}.

Before moving on, we remark that, although we have restricted our attention to stiff epochs due to the fact that PBH formation during a matter-dominated era is not particularly well understood at the time of writing (since the effect of angular momentum and non-sphericity of the collapsing cloud becomes important \cite{Harada:2016mhb,Harada:2017fjm}), the calculation of the GW spectrum presented in this Section is also valid for the case $w=0$, with the only modification being the scalar transfer function in eq.\,(\ref{eq:scalar_trans}), which takes the simpler form $T_\phi(k\eta)=1$ for $\eta<\eta_r$.

\section{Gradual decay into radiation}
\label{sec:gradual}

For the calculation in the previous Section, we assumed that the transition between the reheating stage and the radiation era occurred instantaneously. We now study the case in which the initial fluid decays gradually, thereby making the transition between epochs smooth. This scenario was studied in \cite{Inomata:2019zqy} for the particular case $w=0$ and for a scale-invariant scalar spectrum. Here we extend these results for a generic fluid with constant equation of state parameter $w$ and using a peaked scalar spectrum relevant for PBH formation. It was reported in \cite{Inomata:2019zqy} that, for a scale-invariant scalar spectrum, the induced gravitational wave spectrum in the smooth transition case is suppressed with respect to its sudden counterpart. As we will see, although the same conclusion remains true for a peaked spectrum, the suppression is very mild and therefore does not strongly affect the bounds derived in the previous Section.

In what follows we denote quantities related to the radiation fluid by a subscript $\gamma$, and quantities related to the decaying fluid by a subscript $w$. We model the transition by considering a rate of energy transfer of the form $\Gamma\rho_w$ between the two fluids, with $\Gamma$ constant, such that the total stress-energy tensor $T_{\mu\nu}=T_{\mu\nu}^{\;\;\;(\gamma)}+T_{\mu\nu}^{\;\;\;(w)}$ is conserved, but the separate components are not, so
\begin{equation}
\nabla_\nu T^{\mu\nu(\gamma)}=Q^\mu=-\nabla_\nu T^{\mu\nu(w)}
\end{equation}
for some vector $Q^\mu$ \cite{Malik:2002jb}. The background equations then become, using the number of $e$-folds $dN\equiv\mathcal{H}d\eta$ as the time variable,
\begin{align}
\label{eq:grad_bg1}
3M_p^2H^2&=\rho_\gamma+\rho_w,\\
\frac{d\rho_w}{dN}+3(1+w)\rho_w&=-\frac{\Gamma}{H}\rho_w,\\
\frac{d\rho_\gamma}{dN}+4\rho_\gamma&=\frac{\Gamma}{H}\rho_w.
\label{eq:grad_bg3}
\end{align}

At the perturbative level, we work in Newtonian gauge and assuming no anisotropic stress. We use the following dimensionless set of variables
\begin{equation}
\delta_\gamma\equiv\frac{\delta\rho_\gamma}{\rho_\gamma},\qquad \delta_w\equiv\frac{\delta\rho_w}{\rho_w},\qquad\theta_\gamma\equiv H\frac{\delta q_\gamma}{\rho_\gamma},\qquad\theta_w\equiv H\frac{\delta q_w}{\rho_w},
\end{equation}
where $\delta q_b=a(\rho_b+p_b)\delta v_b$, and $\delta v_b$ denotes the velocity perturbation for each fluid component.\footnote{Note that $\theta$ is sometimes used instead to denote the divergence of the velocity perturbation (e.g.\,in \cite{Inomata:2019zqy}), so one must be careful when comparing different references.} The Einstein equation that determines the evolution of the Newtonian potential $\phi$ is
\begin{equation}
\label{eq:pertsys1}
\frac{d\phi}{dN}+\bigg(1+\frac{1}{3}\frac{k^2}{\mathcal{H}^2}\bigg)\phi+\frac{1}{2}\bigg(\frac{\rho_w}{\rho}\delta_w+\frac{\rho_\gamma}{\rho}\delta_\gamma\bigg)=0.
\end{equation}
The continuity and Euler equations for the fluid are \cite{Malik:2002jb}
\begin{align}
\frac{d\delta_w}{dN}+\frac{\Gamma}{H}\phi-3(1+w)\frac{d\phi}{dN}-\frac{k^2}{\mathcal{H}^2}\theta_w&=0,\\
\frac{d\delta_\gamma}{dN}+\frac{\Gamma}{H}\frac{\rho_m}{\rho_\gamma}(\delta_\gamma-\delta_w-\phi)-4\frac{d\phi}{dN}-\frac{k^2}{\mathcal{H}^2}\theta_\gamma&=0,\\
\frac{d\theta_w}{dN}+\bigg(\frac{3}{2}\frac{\rho+p}{\rho}-\frac{w}{1+w}\frac{\Gamma}{H}-3w\bigg)\theta_w+(1+w)\phi+w\delta_w&=0,\\
\frac{d\theta_\gamma}{dN}+\bigg(\frac{\Gamma}{H}\frac{\rho_w}{\rho_\gamma}+\frac{3p+\rho}{2\rho}\bigg)\theta_\gamma+\frac{4}{3}\phi+\frac{1}{3}\delta_\gamma-\frac{\Gamma}{H}\frac{\rho_w}{\rho_\gamma}\frac{1}{1+w}\theta_w&=0,
\label{eq:pertsys5}
\end{align}
where we have introduced the additional assumption that $\delta p_w=w\delta\rho_w$, and similarly for the radiation component. Together, these five equations form a complete system that can be solved numerically.

\begin{figure}[t]
\begin{center}
$$\includegraphics[width=.44\textwidth]{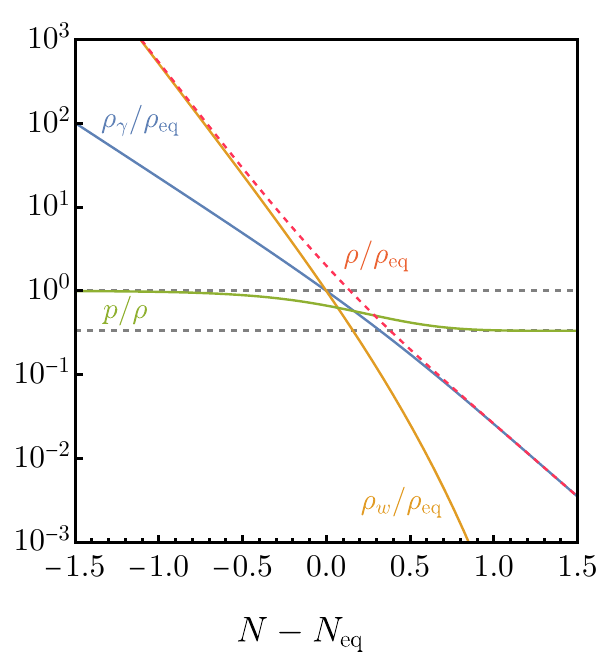}\qquad
\includegraphics[width=.48\textwidth]{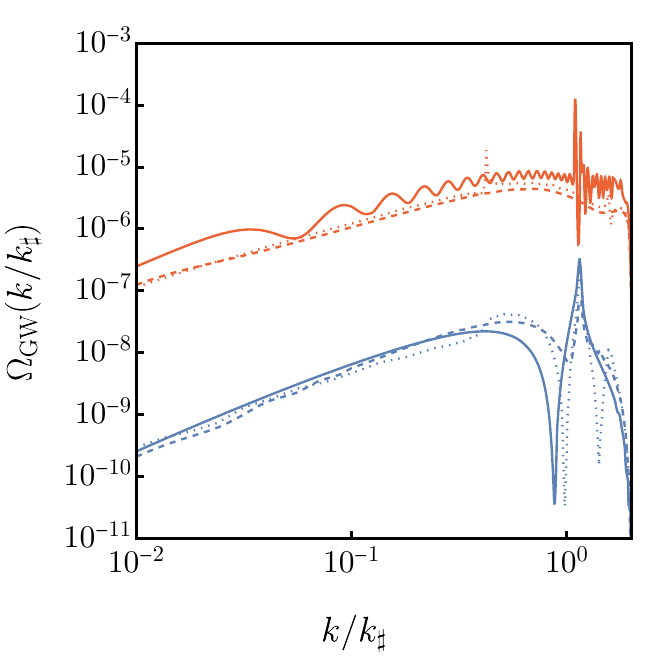}$$
\caption{\em \label{fig:gradual_gws} {
Left panel: evolution of the background energy density of each fluid, as well as the total energy density $\rho$ and the equation of state $p/\rho$, for the case $w=1$. By normalizing all quantities with respect to their value at the time $N_{\rm eq}$ defined by $\rho_\gamma=\rho_w$, the plot becomes independent of $\Gamma$. Right panel: calculation of the GW energy density for the sudden transition case (solid) and the gradual case (dashed) for $w=1$ with $k_\sharp\eta_r=1$ (blue) and $k_\sharp\eta_r=100$ (red). The dotted lines correspond to the gravitational reheating case with $\Gamma=0$. The time $\eta_r$ in the gradual transition scenario is defined through $k_{\rm eq}$ in eq.\,(\ref{eq:keq}). The integrated energy density is very similar for both cases, leaving the bounds of the previous Section essentially unaltered.}}
\end{center}
\end{figure}

To solve the above system of equations, we choose adiabatic initial conditions \cite{Baumann:Cosmo2022}. On superhorizon scales $k\ll\mathcal{H}$ and at sufficiently early times so that $\Gamma\ll H$ and $\rho_\gamma\ll\rho_w$, the equation for the Newtonian potential $\phi$ becomes
\begin{equation}
\frac{d\phi}{dN}+\phi+\frac{1}{2}\delta_w=0.
\end{equation}
Since there is essentially only one fluid present at this time, and we have assumed $\delta p_w=w\delta\rho_w$, the Newtonian potential can be shown to satisfy eq.\,(\ref{eq:newt_pot}) and is therefore frozen on superhorizon scales. Thus, on these scales, we have from the above equation the relation
\begin{equation}
\delta_w=-2\phi.
\end{equation}
The adiabatic initial conditions for the energy density perturbations are \cite{Baumann:Cosmo2022}
\begin{equation}
\frac{\delta_\gamma}{4/3}=\frac{\delta_w}{1+w}.
\end{equation}
Finally, if we assume that $\theta_\gamma$ and $\theta_w$ are also frozen on superhorizon scales, we obtain, from the last two equations,
\begin{equation}
\theta_\gamma=-\frac{8\phi}{9(1+w)},\qquad \theta_w=-\frac{2}{3}\phi.
\end{equation}
By using the relation (\ref{eq:newt_superhorizon}), we can therefore fix all initial conditions in terms of $\mathcal{R}$.

\begin{figure}[t]
\begin{center}
$$\includegraphics[width=.44\textwidth]{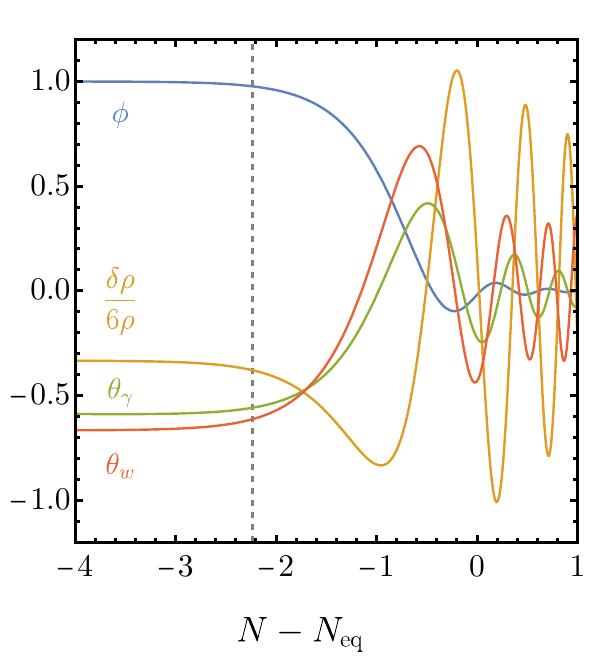}\qquad
\includegraphics[width=.48\textwidth]{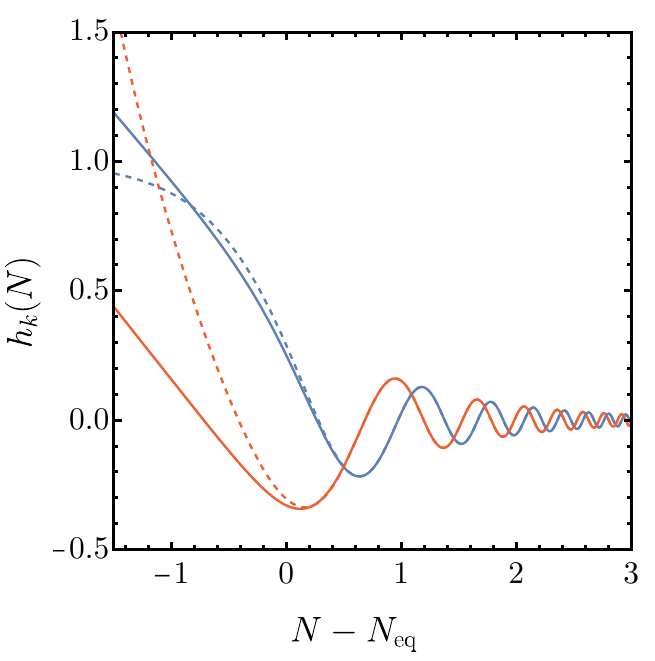}$$
\caption{\em \label{fig:gradual_perts} {
Left panel: evolution of the perturbations with adiabatic initial conditions for $w=0.5$ and $k_\sharp\eta_r=10$, with $\eta_r$ defined via eq.\,(\ref{eq:keq}). The vertical dashed line denotes the horizon crossing time for this mode. Right panel: evolution of the two independent solutions to the homogeneous equation for the tensor modes used to construct the Green's function (solid lines) for $w=1$, $k_\sharp\eta_r=1$, and $k/k_\sharp=1$. At late times the two solutions approach the asymptotic values in eq.\,(\ref{eq:indep_sols}) (dashed lines).
}}
\end{center}
\end{figure}

The evolution of the background energy density of each fluid is shown in Fig.\,\ref{fig:gradual_gws} for $w=1$. Note that, as explained in \cite{Inomata:2019zqy}, the evolution of the background quantities (and in fact, that of the perturbations) is completely independent of the choice of $\Gamma$ once they are normalized by their values at the time $N_{\rm eq}$ defined by $\rho_\gamma(N_{\rm eq})=\rho_w(N_{\rm eq})\equiv \rho_{\rm eq}$, as can be checked from eqs.\,(\ref{eq:grad_bg1}-\ref{eq:grad_bg3}). The left panel of Fig.\,\ref{fig:gradual_perts} shows the time evolution of the perturbations for particular values of the parameters. We define the transition time $\eta_r$ in this case via
\begin{equation}
\label{eq:keq}
\eta_r\equiv\frac{2}{(1+3w)k_{\rm eq}},
\end{equation}
where $k_{\rm eq}$ is the mode that re-enters the horizon at $N_{\rm eq}$. This is the relation between $k_{\rm eq}$ and $\eta_r$ in the sudden transition case of the previous Section, so this definition allows us to make a fair comparison between both scenarios.

\begin{figure}[t]
\begin{center}
$$\includegraphics[width=.48\textwidth]{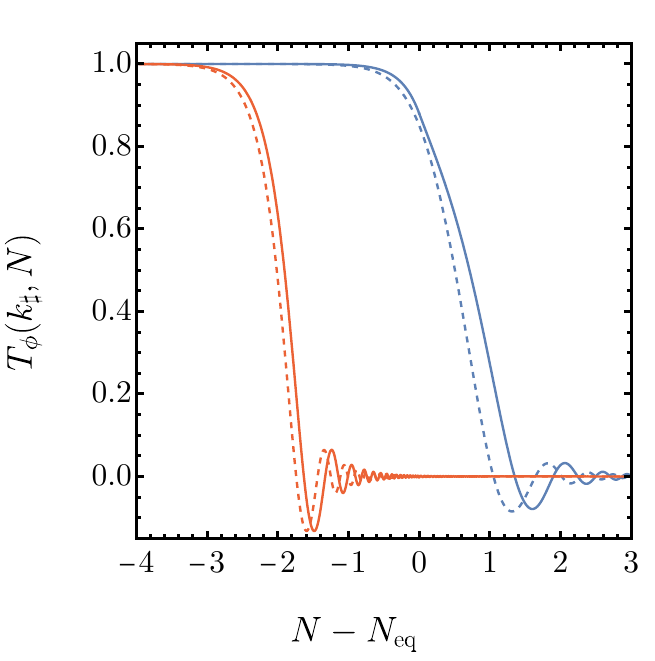}$$
\caption{\em \label{fig:scalar} {
Scalar transfer functions for $w=1$ in the sudden transition case (solid) and the gradual case (dashed) with $k_\sharp\eta_r=1$ (blue) and $k_\sharp\eta_r=100$ (red).
}}
\end{center}
\end{figure}

For the calculation of the gravitational wave spectrum we need both the scalar transfer function $T_\phi$, obtained by solving the system in eqs.\,(\ref{eq:pertsys1}-\ref{eq:pertsys5}) with initial condition $T_\phi=1$ (the condition $\frac{d}{dN}T_\phi=0$ is guaranteed by the choice of adiabatic initial conditions for the other perturbations), and the Green's function for the tensor modes, which we now turn our attention to. The integral in eq.\,(\ref{eq:i_function}) can be rewritten in terms of the number of $e$-folds as
\begin{equation}
\label{eq:ifunc_efolds}
I_k=\bigg(\frac{3+3w}{5+3w}\bigg)^2\int_{-\infty}^\infty \frac{k^2}{\mathcal{H}^2} \frac{h_1(N)h_2(N')-h_2(N)h_1(N')}{h_2(N')\frac{d}{dN}h_1(N')-h_1(N')\frac{d}{dN}h_2(N')} Q(N',p,|{\bm k}-{\bm p}|) dN'.
\end{equation}
The equation of motion for the tensor modes is, in terms of the number of $e$-folds,
\begin{equation}
\label{eq:h_efolds}
\frac{d^2 h^{s}_k}{dN^2}+\frac{3}{2}\bigg(1-\frac{p}{\rho}\bigg)\frac{d h^{s}_k}{dN}+\frac{k^2}{\mathcal{H}^2}h^s_k=0.
\end{equation}
To calculate the GW spectrum today, we must evaluate eq.\,(\ref{eq:ifunc_efolds}) at late times, deep into the radiation era. At this stage the background pressure is $p=\rho/3$, and the Hubble scales as $\mathcal{H}=\mathcal{H}_c(a_c/a)$, where the $c$ subscript denotes some late time $N_c\gg N_{\rm eq}$. The product $a_c\mathcal{H}_c$ approaches a constant value, which can be found numerically, after the transition. Using this scaling in the above equation, we find the following two independent solutions at late times
\begin{equation}
\label{eq:indep_sols}
h_1^{\rm late}= \frac{a_c}{a}\frac{\mathcal{H}_c}{k}\sin\bigg(\frac{a}{a_c}\frac{k}{\mathcal{H}_c}\bigg),\qquad h_2^{\rm late}= \frac{a_c}{a}\frac{\mathcal{H}_c}{k}\cos\bigg(\frac{a}{a_c}\frac{k}{\mathcal{H}_c}\bigg).
\end{equation}
By using these two solutions we can take the time-average of $I_k^2$ as we did in the previous Section, and we obtain, after performing the momentum integral over the Dirac delta power spectrum,
\begin{equation}
\frac{k^2}{\mathcal{H}^2}\langle I_k^2\rangle=\frac{1}{2}(J_s^2+J_c^2)=\frac{1}{2}J^2\bigg(\frac{k}{k_{\rm eq}},\frac{k_\sharp}{k_{\rm eq}}\bigg),
\end{equation}
where
\begin{align}
J_s&=\bigg(\frac{3+3w}{5+3w}\bigg)^2\int_{-\infty}^\infty \frac{(k/k_{\rm eq})^2}{(\mathcal{H}/k_{\rm eq})^2} G_2(N',k/k_{\rm eq}) Q(N',k_\sharp/k_{\rm eq}) dN',\\
J_c&=\bigg(\frac{3+3w}{5+3w}\bigg)^2\int_{-\infty}^\infty \frac{(k/k_{\rm eq})^2}{(\mathcal{H}/k_{\rm eq})^2}  G_1(N',k/k_{\rm eq}) Q(N',k_\sharp/k_{\rm eq}) dN',
\end{align}
with
\begin{align}
\label{eq:gfunc1}
G_1(N',k/k_{\rm eq})&\equiv\frac{h_1(N')}{h_2(N')\frac{d}{dN}h_1(N')-h_1(N')\frac{d}{dN}h_2(N')},\\
G_2(N',k/k_{\rm eq})&\equiv\frac{h_2(N')}{h_2(N')\frac{d}{dN}h_1(N')-h_1(N')\frac{d}{dN}h_2(N')}.
\label{eq:gfunc2}
\end{align}
Once again we see that the GW energy depends only on $k/k_{\rm eq}=(k/k_\sharp)(k_\sharp/k_{\rm eq})$, as well as the two parameters $k_\sharp/k_{\rm eq}$ and $w$. The dependence on the parameter $k_{\rm eq}$ therefore replaces $\eta_r$ from the previous section, with the relation between both given by eq.\,(\ref{eq:keq}).

We remark that, although the Green's function does not depend on which two linearly independent solutions are chosen to construct it, the fact that we perform the time average over $I_k^2$ by using $\langle\sin^2(k\eta)\rangle=\langle\cos^2(k\eta)\rangle=1/2$ means we need to project the two solutions that behave as eq.\,(\ref{eq:indep_sols}) at late times. This can be accomplished by imposing the boundary conditions $h_{1,2}(N_{\rm late})=h_{1,2}^{\rm late}(N_{\rm late})$ at some late time $N_{\rm late}\gg N_{\rm eq}$, sufficiently deep into the radiation era (numerically, we find that a few $e$-folds after the transition suffice). The corresponding solutions are shown in the right panel of Fig.\,\ref{fig:gradual_perts} together with their late time limits for $w=1$, $k_\sharp\eta_r=1$ and $k/k_\sharp=1$.

We show the resulting GW energy density for $w=1$ in the right panel of Fig.\,\ref{fig:gradual_gws}, for two illustrative examples with $k_\sharp\eta_r=1$ and $k_\sharp\eta_r=100$. We find that, although the signal is very mildly suppressed with respect to the sudden transition case of the previous Section, the difference between both is essentially negligible for the purpose of estimating the curve in Fig.\,\ref{fig:gw_constraints}, so we conclude that the results from the previous Section are robust. We nevertheless remark that this conclusion might change if the transition is modelled differently. In the same panel, we show the gravitational reheating case with $\Gamma=0$. We find that in this case the oscillations are also washed out, and the amplitude of the signal is comparable to the gradual transition case. The transfer functions obtained numerically using the procedure in this Section are compared to the results using eq.\,(\ref{eq:scalar_trans}) in Fig.\,\ref{fig:scalar}.

Having established that the gradual transition scenario is not significantly different from the sudden case, we go back to using the analytical formulas of the latter in the following Section.

\section{Smoothed lepton number density}
\label{sec:smooth_lepton}

In this Section we estimate the size of the baryon asymmetry fluctuations induced by the gravitational chiral anomaly. The quantity of interest is the variance of the lepton number density\footnote{Throughout this Section we often abuse language and refer to this quantity simply as the baryon asymmetry. In the leptogenesis scenario, the relation between the baryon and lepton number densities is, assuming the Standard Model matter content, $n_{\rm L}=(79/28)n_{\rm B}$ \cite{Weinberg:2008zzc}. The measured value for this quantity is $n_{\rm L}/s\simeq 2\times 10^{-10}$ \cite{Planck:2018vyg,Mossa:2020gjc}.} $n_{\rm L}$ computed in eqs.\,(\ref{eq:nlo_fourier}) and (\ref{eq:nnlo_ori}), averaged over a region of size $r_\sigma$,
\begin{equation}
\label{eq:smoothed_nl}
\big\langle |n_{\rm L}|^2\big\rangle_{r_\sigma} \equiv \bigg\langle\bigg|\int d^3{\bm r}W_{r_\sigma}(r)n_{\rm L}({\bm x}+{\bm r})\bigg|^2\bigg\rangle,
\end{equation}
where $W_{r_\sigma}(r)$ is a Window function that decays smoothly on scales $r\gg r_\sigma$, which we take to be a Gaussian for concreteness,
\begin{equation}
\label{eq:window_gauss}
W_{r_\sigma}(r)=\bigg(\frac{1}{r_\sigma\sqrt{\pi}}\bigg)^3e^{-r^2/r_\sigma^2}.
\end{equation}

Let us clarify why we focus on the variance of this quantity. We would naively expect the baryon asymmetry to be determined by the mean value of $n_{\rm L}$. However, in the absence of a chiral gravitational wave background (generated for instance, via some inflationary coupling of the form $f(\varphi) R\tilde R$ such as the one considered in \cite{Alexander:2004us}, where $\varphi$ is the inflaton field), which would make the terms quadratic in $h_{ij}$ of the form shown in eq.\,(\ref{eq:tensor_squared}) nonzero, the mean value $\langle n_{\rm L}\rangle$ vanishes. This can be easily seen at NLO, since, schematically, taking the mean value of eq.\,(\ref{eq:nlo_fourier}) yields $\langle\phi h\rangle\sim \langle\phi\rangle\langle h\rangle=0$. The fact that the mean value also vanishes at NNLO is less obvious. The dimensionless bispectrum $\mathcal{B}_\mathcal{R}$ of curvature perturbations is defined by
\begin{equation}
\big\langle \mathcal{R}_{\bm q}\mathcal{R}_{\bm k}\mathcal{R}_{\bm{p}-\bm{k}}\big\rangle\equiv(2\pi)^3\delta^3_{\bm{p}+\bm{q}}\frac{(2\pi^2)^2}{q^2k^2|\bm{p}-\bm{k}|^2}\mathcal{B}_\mathcal{R}(q,k,|\bm{p}-\bm{k}|),
\end{equation}
where we have introduced the shorthand $\delta^3_{\bm{p}}\equiv\delta^3(\bm{p})$ for later convenience. This quantity vanishes if the fluctuations are Gaussian and is therefore slow-roll suppressed in conventional inflationary scenarios, but this could change depending on the dynamics of the inflaton (particularly in models of PBH production featuring a potential with a near-inflection point, in which the slow-roll approximation breaks down \cite{Taoso:2021uvl}). After some manipulation, we obtain, from eq.\,(\ref{eq:nnlo_ori}),
\begin{align}
a^3\big\langle |n_{\rm L}|\big\rangle_{r_\sigma}&=\frac{\pi^2}{8}\bigg(\frac{3+3w}{5+3w}\bigg)\sum_s\int d\eta\int\frac{d^3{\bm p}}{(2\pi)^3}\epsilon_{ijk}(p_j p_\ell p_k)e_{i\ell}^s({\bm p})T_\phi(p\eta)\nonumber\\
&\quad
\int\frac{d^3{\bm k}}{(2\pi)^3}\Big[{\bm k}\cdot {\bm e}^s({\bm p})\cdot {\bm k}\Big]I_p'(\eta,k,|{\bm p}-{\bm k}|)\frac{\mathcal{B}_\mathcal{R}(p,k,|{\bm p}-{\bm k}|)}{p^4k^2|{\bm p}-{\bm k}|^2},
\label{eq:nnlo_meanvalue}
\end{align}
where we have used the Dirac delta function in the definition of the bispectrum to perform one of the momentum integrals, as well as the normalization of the window function
\begin{equation}
\int d^3{\bm r} W_{r_\sigma}(r)=1.
\end{equation}
The right-hand side of eq.\,(\ref{eq:nnlo_meanvalue}) clearly vanishes, since $p_jp_k$ is symmetric, but $\epsilon_{ijk}$ is not. In what follows we assume that the curvature perturbation $\mathcal{R}$ follows a Gaussian distribution.

Instead of focusing on the mean value of this quantity, we assume that the baryon asymmetry of the Universe is generated via some other mechanism which we remain agnostic about, and shift our attention to the variance of $n_{\rm L}$.\footnote{It is conceivable that the mechanism responsible for generating the baryon asymmetry will also lead to a spectrum of fluctuations which could be potentially larger than the contributions discussed here. The relative importance of each term is of course model-dependent and can only be assessed once a particular mechanism is fixed.} As noted in \cite{Maroto:2022xrv}, since inflation is a stochastic process, we expect the lepton number density to deviate from its mean value in different patches. This deviation will generically have a magnitude of order $\sim\sqrt{\langle |n_{\rm L}|^2\rangle_{r_\sigma}}$ in a region of size $r_\sigma$, leading to fluctuations in the baryon asymmetry. Before moving on with the calculation however, let us note that, as discussed in \cite{Maroto:2022xrv}, this asymmetry does not survive at late times on arbitrarily small patches, due to annihilation processes. Let us consider two neighboring patches, one with a matter excess, and another one with an antimatter excess. If particles are able to freely travel from one patch to the other, annihilation will take place, leading to a smaller asymmetry overall. On sufficiently large patches, however, the asymmetry always survives at late times. To see why, note that the maximum distance a particle in the radiation bath can travel between collisions is $\lambda/a$, where $\lambda$ is its mean free path. In a time $\Delta t$, the particle undergoes $N=\Delta t/\lambda$ collisions. The average displacement in a time $\Delta t$ for a random walk is therefore $\Delta r=(\lambda/a)\sqrt{N}=(\lambda/a)\sqrt{\Delta t/\lambda}$. Integrating this displacement yields the Silk length \cite{Kolb:1990vq}, which sets the limit below which annihilation can take place.\footnote{This is analogous to photon diffusion in the CMB \cite{Baumann:Cosmo2022}.} On the other hand, at late times, after the electroweak sphaleron processes have taken place, the asymmetry is carried by quarks, which are relativistic. Once the temperature is low enough, quarks are confined into non-relativistic baryons and their mean free path drops significantly. The Silk length is therefore approximately given by
\begin{equation}
r_{\rm S}^2\simeq \int_0^{\rm t_{\rm QCD}}dt\;\frac{\lambda_{\rm Q}}{a^2},
\end{equation}
where $\lambda_{\rm Q}=1/\Gamma_{\rm Q}$ denotes the mean free path of quarks in the plasma, and $\Gamma_{\rm Q}$ is the interaction rate, which can be estimated as $\Gamma_{\rm Q}\simeq T$ \cite{Thoma:1993vs}. For simplicity, let us neglect the effect of the equation of state on the evolution of $a$ and assume that the Universe is always radiation-dominated. We also neglect the change in the relativistic degrees of freedom in entropy $g_{\star s}$ between the end of inflation and the QCD scale. We then have $T\simeq T_0/a$ and $H=H_0a^{-2}$, which leads to
\begin{equation}
r_{\rm S}^2\simeq \int_0^{a_{\rm QCD}}da\;\frac{\Gamma_{\rm Q}^{-1}}{a^3H}\simeq \frac{1}{H_0T_{\rm QCD}}\simeq \Big(10^{-17}{\rm Mpc}\Big)^{2}.
\end{equation}
We remark that this is only an order of magnitude estimate,\footnote{A similar value of $r_{\rm S}\simeq 10^{-16}{\rm Mpc}$ was assumed in \cite{Maroto:2022xrv}. However, the specific number is not particularly important for our purposes since, as we will see, the peak in the spectrum of fluctuations generated by PBHs lies far above this scale.} and an accurate description of the damping dynamics requires solving the Boltzmann equation. We conclude that the asymmetry is preserved on scales larger than the Silk length of quarks at the confinement scale, $r_\sigma \gtrsim r_{\rm S} \simeq 10^{-17}{\rm Mpc}$. This is the domain of validity of the calculations in this Section. Below this scale, annihilation dynamics become important and we expect the amplitude of the baryon asymmetry fluctuations to drop significantly.

\begin{figure}[t]
\begin{center}
$$\includegraphics[width=0.75\textwidth]{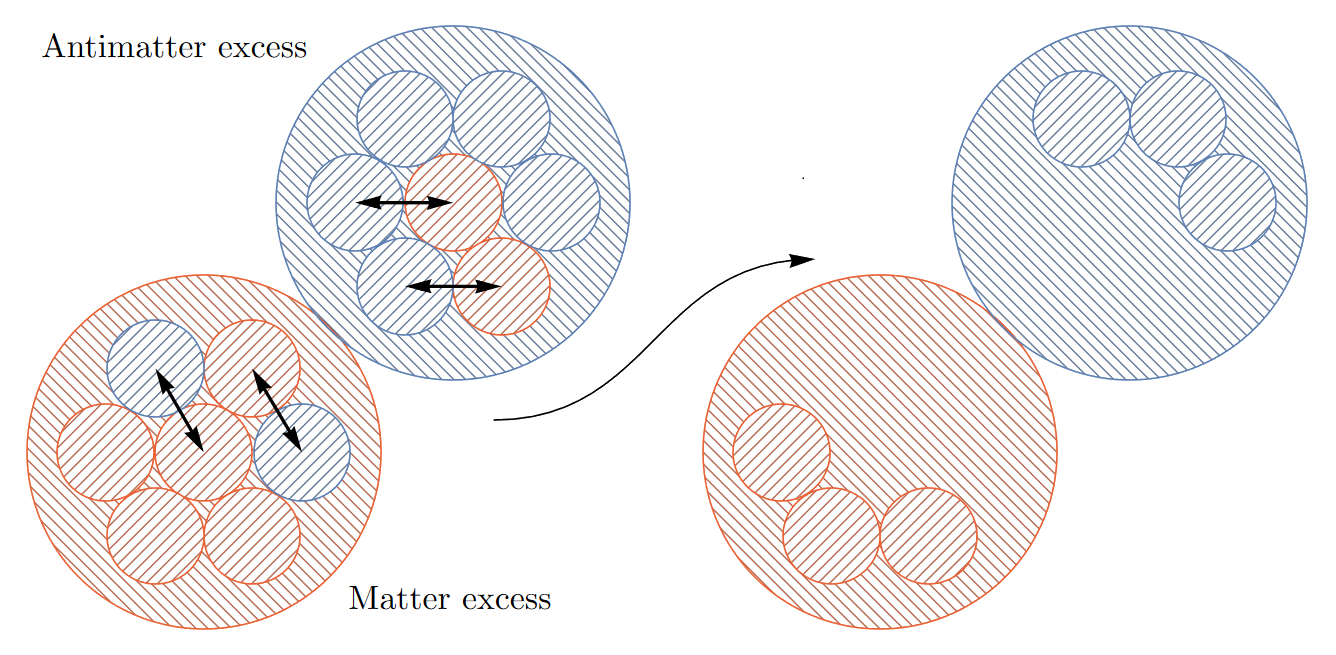}$$
\caption{\em \label{fig:schematic} {
Schematic depiction of the late-time behaviour of the baryon asymmetry fluctuations. Small neighbouring patches with matter and antimatter excesses annihilate, homogenizing the distribution on small scales. Larger regions remain unaffected, since quarks cannot travel from one patch to the other if the distance separating them is larger than the Silk length (see text). This length scale sets the range of validity for the calculations in this Section.
}}
\end{center}
\end{figure}

In what follows we will compute the fluctuations in three different cases, 1.\,for the NLO term in eq.\,(\ref{eq:nlo_fourier}) with a flat scalar spectrum (\ref{eq:pr_flat}) and with the tensor spectrum given by the relation $\mathcal{P}_h=r\mathcal{A}_\flat$, where $r\lesssim 10^{-2}$ \cite{BICEP:2021xfz,Tristram:2021tvh} is the tensor-to-scalar ratio, 2.\,the same NLO term with a sharp scalar spectrum (\ref{eq:pr_sharp}), but with a flat tensor spectrum given by the same relation as in the first case, and 3.\,the NNLO term in eq.\,(\ref{eq:nnlo_ori}) for a peaked scalar spectrum, and with the tensor modes induced by scalar perturbations. As we will see, each term dominates at different scales, and the total asymmetry fluctuations will in general be given by the sum of the three contributions, plus the subdominant mixed terms that we have neglected as per the arguments of Section \ref{sec:lepton_anomaly}. We also neglect the term involving the flat part of the scalar spectrum together with the induced GW piece, since we expect it to be suppressed with respect to the NNLO term mentioned above. Before moving on with the technical details of the calculation, let us anticipate that, since the variance of $n_{\rm L}$ is proportional to both the tensor and scalar power spectrum, from dimensional analysis we can guess the result in all three cases to be $\big\langle |n_{\rm L}|^2\big\rangle_{r_\sigma}^{1/2}\sim \sqrt{\mathcal{P}_\mathcal{R}\mathcal{P}_h}/(a r_\sigma)^3$, since $r_\sigma$ is the only dimensionful parameter. Dividing by the entropy density $s \sim g_{\star s}(T)T^3$ and using entropy conservation, we find
\begin{equation}
\frac{n_{\rm L}}{s}\sim \frac{\sqrt{\mathcal{P}_\mathcal{R}\mathcal{P}_h}}{g_{\star s}(T_0)(r_\sigma T_0)^3},
\end{equation}
up to an overall numerical factor. The NNLO term will be highly peaked around $r_\sigma \sim k_\sharp^{-1}$, whereas the NLO term due to the flat part of the power spectrum will instead grow simply as $r_\sigma^{-3}$ and reach its highest value at the Silk scale $r_{\rm S}$, as per the previous discussion.

\bigskip
\noindent{\bf Case 1: NLO, flat scalar spectrum (NLO$\flat$)}

\noindent Putting together eqs.\,(\ref{eq:nlo_fourier}) and (\ref{eq:smoothed_nl}), we find, after some manipulation,
\begin{align*}
\big\langle a^6 |n_{\rm L}|^2\big\rangle_{r_\sigma}&=\bigg(\frac{1}{16\pi^2}\bigg)^2\sum_{st}\int d\eta\int d\hat{\eta}\int\frac{d^3 {\bm p}}{(2\pi)^3}\int\frac{d^3 {\bm q}}{(2\pi)^3}\int\frac{d^3 \hat{{\bm p}}}{(2\pi)^3}\int\frac{d^3 \hat{{\bm q}}}{(2\pi)^3}\int d^3{\bm r}\int d^3\hat{{\bm r}}\\
&\quad \bigg(\frac{3+3w}{5+3w}\bigg)^2 T_\phi(q\eta)T_\phi(\hat{q}\hat{\eta})T_h^{\prime}(p\eta)T_h^{\prime}(\hat{p}\hat{\eta})
\big\langle\mathcal{R}_{\bm q}\mathcal{R}_{\hat{\bm q}}^\dagger\big\rangle\big\langle h_{\bm p}^{s}(0)h_{\hat{\bm p}}^{t\dagger}(0)\big\rangle \\
&\quad W_{r_\sigma}(r)W_{r_\sigma}(\hat{r}) e^{i({\bm p}+{\bm q})\cdot ({\bm x}+{\bm r})}e^{-i(\hat{{\bm p}}+\hat{{\bm q}})\cdot ({\bm x}+\hat{{\bm r}})}\Big[\epsilon_{ijk}\epsilon_{abc} q_j q_\ell p_k e_{i\ell}^s({\bm p}) \hat{q}_b \hat{q}_d \hat{p}_c e_{ad}^t(\hat{{\bm p}}) \Big].\numberthis
\end{align*}
Straightforward evaluation of the correlation functions yields
\begin{align}
\big\langle\mathcal{R}_{\bm k}\mathcal{R}_{\bm p}^\dagger\big\rangle&=\frac{2\pi^2}{k^3}\mathcal{P}_\mathcal{R}(k)\delta^3_{{\bm k}-{\bm p}}(2\pi)^3,\\
\big\langle h_{\bm k}^s(0)h_{\bm p}^{t\dagger}(0)\big\rangle&=\frac{2\pi^2}{k^3}\mathcal{P}_h(k)\delta^3_{{\bm k}-{\bm p}}(2\pi)^3\delta^{st},
\end{align}
where we have introduced the shorthand $\delta^3({\bm k})\equiv \delta^3_{\bm k}$ and assumed that $h_k^+=h_k^\times=h_k$. Using the Dirac delta functions to perform two of the momentum integrals, we find
\begin{align*}
\big\langle a^6 |n_{\rm L}|^2\big\rangle_{r_\sigma}&=\frac{1}{64}\bigg(\frac{3+3w}{5+3w}\bigg)^2
\int d\eta\int d\hat{\eta}\int\frac{d^3 {\bm p}}{(2\pi)^3}\int\frac{d^3 {\bm q}}{(2\pi)^3}\frac{\mathcal{P}_\mathcal{R}(q)}{q^3}
\frac{\mathcal{P}_h(p)}{p^3}\big|\widehat{W}_{r_\sigma}(|{\bm p}+{\bm q}|)\big|^2\\
&\quad T_\phi(q\eta)T_\phi(q\hat{\eta})T_h^{\prime}(p\eta)T_h^{\prime}(p\hat{\eta})\bigg[\sum_{s}\epsilon_{ijk}\epsilon_{abc} q_j q_\ell p_k e_{i\ell}^s({\bm p}) q_b q_d p_c e_{ad}^s({\bm p}) \bigg],\numberthis
\end{align*}
where we have introduced the Fourier transform of the window function,
\begin{equation}
\widehat{W}_{r_\sigma}(k)\equiv\int d^3{\bm r}\;W_{r_\sigma}(r) e^{-i{\bm k}\cdot{\bm r}}.
\end{equation}

By expanding the Levi-Civita symbols in terms of Kronecker deltas, we find the following expression for the term in brackets
\begin{equation}
\sum_s\epsilon_{ijk}\epsilon_{abc} q_j q_\ell p_k e_{i\ell}^s({\bm p}) q_b q_d p_c e_{ad}^s({\bm p})=\Big[q^2p^2-({\bm q}\cdot{\bm p})^2\Big]\Big[{\bm q}\cdot {\bm e}^s({\bm p})\cdot {\bm e}^s({\bm p})\cdot{\bm q}\Big]-p^2\Big[{\bm q}\cdot {\bm e}^s({\bm p})\cdot{\bm q}\Big]^2.
\end{equation}
If we choose the coordinate system in such a way that the ${\bm z}$ axis is aligned with ${\bm p}$, then the vectors in the definition of the polarization tensors in eq.\,(\ref{eq:pol_tensors}) are simply ${\bm v}={\bm x}$ and $\bar{{\bm v}}={\bm y}$. Using $\theta$ to denote the angle between ${\bm q}$ and the ${\bm z}$ axis, and $\varphi$ for the corresponding azimuthal angle, we obtain, after some straightforward algebra,
\begin{equation}
\sum_s\epsilon_{ijk}\epsilon_{abc} q_j q_\ell p_k e_{i\ell}^s({\bm p}) q_b q_d p_c e_{ad}^s({\bm p})=\frac{1}{2}p^2q^4\sin^4\theta.
\end{equation}

The window function is obtained by Fourier-transforming eq.\,(\ref{eq:window_gauss}), \begin{equation}
\widehat{W}_{r_\sigma}(|{\bm p}+{\bm q}|)=\exp\bigg[-\frac{r_\sigma^2}{4}(p^2+q^2+2pq\cos\theta)\bigg].
\end{equation}

After switching to spherical coordinates, all of the angular integrals can be performed explicitly, and we arrive at the expression
\begin{align}
\big\langle a^6 |n_{\rm L}|^2\big\rangle_{r_\sigma}&=\frac{1}{64\pi^4}\bigg(\frac{3+3w}{5+3w}\bigg)^2\int dp\int dq\;pq^3\mathcal{P}_\mathcal{R}(q)
\mathcal{P}_h(p)\bigg|\int d\eta\; T_\phi(q\eta)T_h^{\prime}(p\eta)\bigg|^2\nonumber
\\
&\quad
\exp\bigg[-\frac{r_\sigma^2}{2}(p^2+q^2)\bigg]
\bigg[-3\frac{\cosh(pqr_\sigma^2)}{(pqr_\sigma^2)^4}+(3+p^2q^2r_\sigma^4)\frac{\sinh(pqr_\sigma^2)}{(pqr_\sigma^2)^5}\bigg].
\label{eq:nlob_fourier}
\end{align}

\begin{figure}[t]
\begin{center}
$$\includegraphics[width=.45\textwidth]{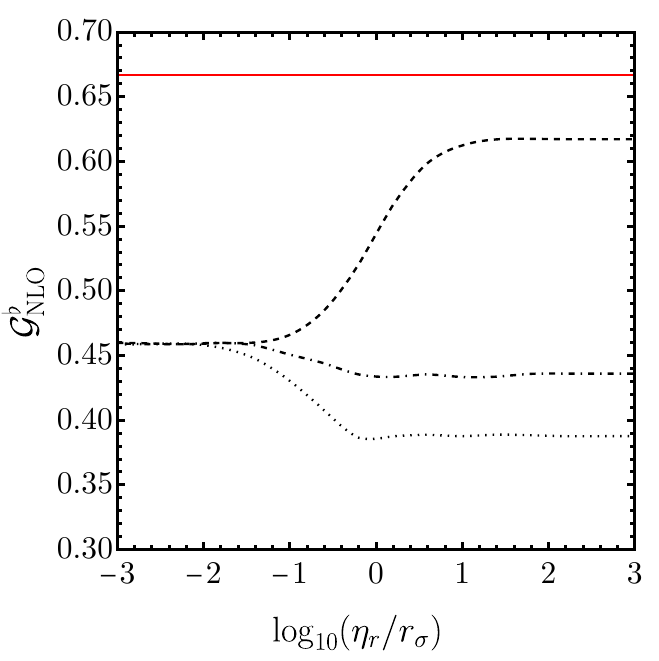}$$
\caption{\em \label{fig:gFlatNLO} {
Numerical calculation of the coefficient $\mathcal{G}_{\rm NLO}^\flat$ in eq.\,(\ref{eq:gflat_int}) as a function of the parameter $\eta_r/r_\sigma$ for the different equations of state $w=0$ (black, dashed), $w=0.5$ (black, dot-dashed), and $w=1$ (black, dotted), together with the analytical estimate in eq.\,(\ref{eq:gflat_approx}) (red, solid).}}
\end{center}
\end{figure}

As explained earlier, the observable quantity of interest is the root mean square of the lepton number density. In order to compare with the baryon asymmetry after the electroweak sphaleron processes have taken place, we need to divide by the entropy density
\begin{equation}
s=\frac{2\pi^2}{45}g_{\star s}(T)T^3.
\end{equation}
The main contribution to the time integrals in eq.\,(\ref{eq:nlob_fourier}) occurs during the reheating stage, since the transfer functions for both tensors and scalars decay quickly during the radiation era. Let us suppose that the integrals freeze at some time $t_f$ shortly after reheating ends. Then, all of the time dependence of the lepton number density is in the $a^3$ factor, and we can relate the value of $n_{\rm L}/s$ to its value today by using entropy conservation,
\begin{equation}
g_{\star s}(T_f)a_f^3T_f^3=g_{\star s}(T_0)a_0^3T_0^3.
\end{equation}
Moreover, the upper limit in the time integrals can then be taken as $\eta\rightarrow\infty$.

So far, we have not made any explicit choice for the power spectra. Let us set
\begin{equation}
\label{eq:flat_spectra}
\mathcal{P}_\mathcal{R}(k)=\mathcal{A}_\flat,\qquad \mathcal{P}_h(k)=r\mathcal{A}_\flat.
\end{equation}
We then obtain
\begin{equation}
\frac{n_{\rm L}}{s}\bigg|_{\rm NLO\flat}=\frac{45}{16\pi^4}\bigg(\frac{3+3w}{5+3w}\bigg)\frac{\sqrt{r\mathcal{A}_\flat^2\mathcal{G}^\flat_{\rm NLO}}}{g_{\star s}(T_0)(T_0r_\sigma)^3},
\end{equation}
with
\begin{align}
\mathcal{G}^\flat_{\rm NLO}&=\int_0^\infty dp_\sigma\int_0^\infty dq_\sigma\; \bigg|\int_0^\infty dx\;
T_\phi\bigg(\frac{q_\sigma}{p_\sigma}x,q_\sigma\frac{\eta_r}{r_\sigma}\bigg)\frac{d}{dx}T_h\bigg(x,p_\sigma\frac{\eta_r}{r_\sigma}\bigg)\bigg|^2\nonumber
\\
&\quad
\exp\bigg[-\frac{1}{2}(p_\sigma^2+q_\sigma^2)\bigg]
\bigg[-3\frac{\cosh(p_\sigma q_\sigma)}{p_\sigma^3 q_\sigma}+(3+p_\sigma^2q_\sigma^2)\frac{\sinh(p_\sigma q_\sigma)}{p_\sigma^4 q_\sigma^2}\bigg],
\label{eq:gflat_int}
\end{align}
where we have introduced the dimensionless variables
\begin{equation}
x=p\eta,\qquad p_\sigma=r_\sigma p,\qquad q_\sigma=r_\sigma q,
\end{equation}
as well as the following notation for the transfer functions
\begin{equation}
T_\phi(k\eta)\equiv T_\phi(k\eta,k\eta_r),\qquad T_h(k\eta)\equiv T_h(k\eta,k\eta_r),
\end{equation}
which simply makes explicit the dependence on $\eta_r$. There is, of course, also an implicit dependence on the equation of state $w$ during the reheating stage. In fact, $\mathcal{G}^\flat_{\rm NLO}$ is a dimensionless numerical factor that depends only on the parameters $w$ and $\eta_r/r_\sigma$, which can be found from eq.\,(\ref{eq:eta_temp}),
\begin{equation}
\frac{\eta_r}{r_\sigma}\simeq\bigg(\frac{1}{1+3w}\bigg)\bigg(\frac{r_\sigma^{-1}}{10^{14}{\rm Mpc}^{-1}}\bigg)\bigg(\frac{10^{7}{\rm GeV}}{T_r}\bigg).
\end{equation}
The dependence on both of these quantities is quite mild, as illustrated by the numerical results presented in Fig.\,\ref{fig:gFlatNLO}.

It is instructive to obtain an analytical result for $\mathcal{G}^\flat_{\rm NLO}$ by introducing some approximations, following the procedure in \cite{Maroto:2022xrv}. The first is that since the time integrals freeze around the end of reheating, we can take the upper limit as $\eta_r$ instead of $\infty$, which allows us to use the expressions for the transfer functions valid during the reheating stage. Moreover, since the dependence on the equation of state $w$ is quite mild, we can simply set $w=0$, so that $T_\phi(k\eta)=1$. The time integral can then be performed immediately, yielding
\begin{equation}
\int_0^{p\eta_r} dx\;
T_\phi\bigg(\frac{q_\sigma}{p_\sigma}x\bigg)\frac{d}{dx}T_h(x)=T_h\bigg(p_\sigma\frac{\eta_r}{r_\sigma}\bigg)-1,
\end{equation}
where we have suppressed the second arguments in the transfer functions, since we are using the expressions during the reheating stage, and no matching of coefficients is involved. We can finally make the assumption that $\eta_r\gg r_\sigma$, so that $T_h(p_\sigma\eta_r/r_\sigma)\rightarrow 0$. The remaining integrals over $p_\sigma$ and $q_\sigma$ can then be performed analytically, and we find
\begin{equation}
\label{eq:gflat_approx}
\mathcal{G}^\flat_{\rm NLO}=\frac{2}{3}.
\end{equation}
This result is very close to those obtained by numerically calculating the integral in eq.\,(\ref{eq:gflat_int}), as shown in Fig.\,\ref{fig:gFlatNLO}, even for $\eta_r\lesssim r_\sigma$ and $w\neq 0$.

Our result differs from the one in \cite{Maroto:2022xrv} in two ways. The first is that we obtain a different expression for the integral in eq.\,(\ref{eq:gflat_int}). The momentum integrals in \cite{Maroto:2022xrv} diverge in both the IR and UV, so the authors have calculated it by imposing cutoffs on both limits. Our integral, in contrast, converges without the need for cutoffs, so we obtain a finite result for all parameter choices. The second difference is that the evolution of the transfer functions during the radiation era was not considered in \cite{Maroto:2022xrv}, and instead an analytical estimate similar to the one performed above was presented. We have taken the effect of this evolution into account in our calculation, and computed the integrals in eq.\,(\ref{eq:gflat_int}) numerically by varying the two parameters $\eta_r/r_\sigma$ and $w$, confirming that the dependence of the result on these quantities is very mild.

\begin{figure}[t]
\begin{center}
$$\includegraphics[width=.44\textwidth]{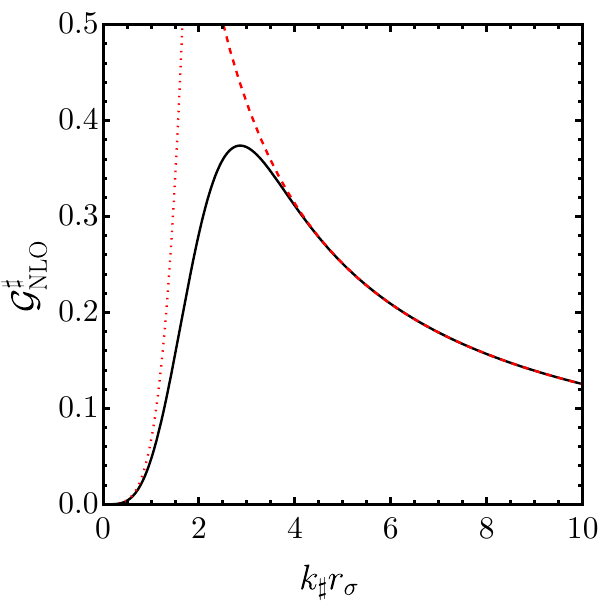}\qquad
\includegraphics[width=.45\textwidth]{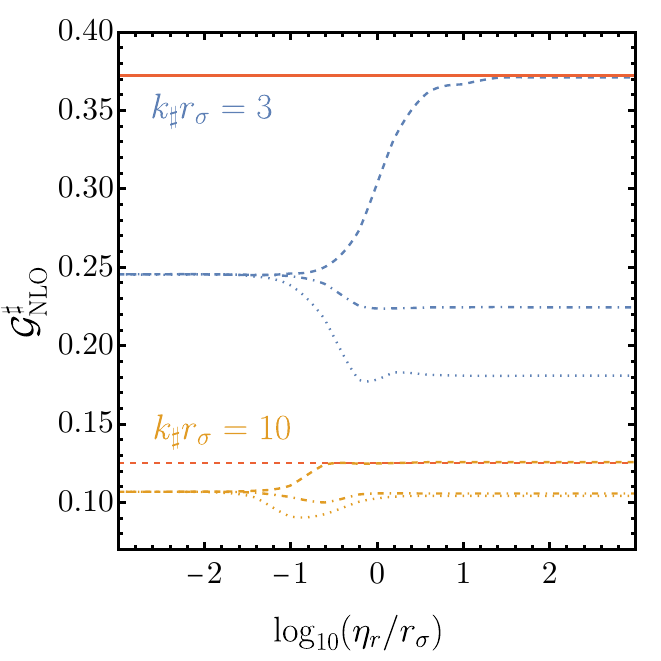}
$$
\caption{\em \label{fig:gsharp_approx} {
Left panel: approximate expression for $\mathcal{G}^\sharp_{\rm NLO}$ in eq.\,(\ref{eq:gsharp_approx}) (solid black), together with the asymptotic limits in (\ref{eq:asymp_nlo}) (dashed and dotted red). Right panel: numerical calculation of the full integral in eq.\,(\ref{eq:nlo_sharp_fullint}) for $k_\sharp r_\sigma=3$ (blue) and $k_\sharp r_\sigma=10$ (orange) for $w=0$ (dashed), $w=0.5$ (dot-dashed), and $w=1$ (dotted), together with the approximate expression depicted in the left panel for each case (solid red and dashed red).
}}
\end{center}
\end{figure}

We conclude that, apart from the mild dependence of $\mathcal{G}^\flat_{\rm NLO}$ on $r_\sigma$, the resulting $n_{\rm L}/s$ scales as $r_\sigma^{-3}$. As discussed earlier, this result is only valid up to the Silk scale $r_{\rm S}$, where the annihilation damping becomes relevant. The maximum value it reaches is therefore, taking $r=0.01$ \cite{BICEP:2021xfz,Tristram:2021tvh} and $\mathcal{G}^\flat_{\rm NLO}=2/3$ for simplicity,
\begin{equation}
\frac{n_{\rm L}}{s}\bigg|_{\rm NLO\flat}\simeq \frac{45}{16\pi^4}\bigg(\frac{3+3w}{5+3w}\bigg)\frac{\sqrt{r\mathcal{A}_\flat^2(2/3)}}{g_{\star s}(T_0)(T_0r_{\rm S})^3}\simeq 10^{-38}.
\end{equation}
The asymmetry over the entire Hubble patch today is much smaller,
\begin{equation}
\frac{n_{\rm L}}{s}\bigg|_{\rm NLO\flat}\simeq \frac{45}{16\pi^4}\bigg(\frac{3+3w}{5+3w}\bigg)\frac{\sqrt{r\mathcal{A}_\flat^2(2/3)}}{g_{\star s}(T_0)}\bigg(\frac{H_0}{T_0}\bigg)^3\simeq 10^{-100},
\label{eq:hubble_est}
\end{equation}
so we confirm, as claimed in \cite{Maroto:2022xrv}, that these fluctuations alone cannot explain the observed value of $n_{\rm L}/s\simeq 10^{-10}$. The same conclusion holds for the two following cases, but, as we will see momentarily, the resulting spectrum of fluctuations has a rich structure that could, in principle, allow us to probe different inflationary models if it were observable on small scales. 

\begin{figure}[t]
\begin{center}
$$\includegraphics[width=.45\textwidth]{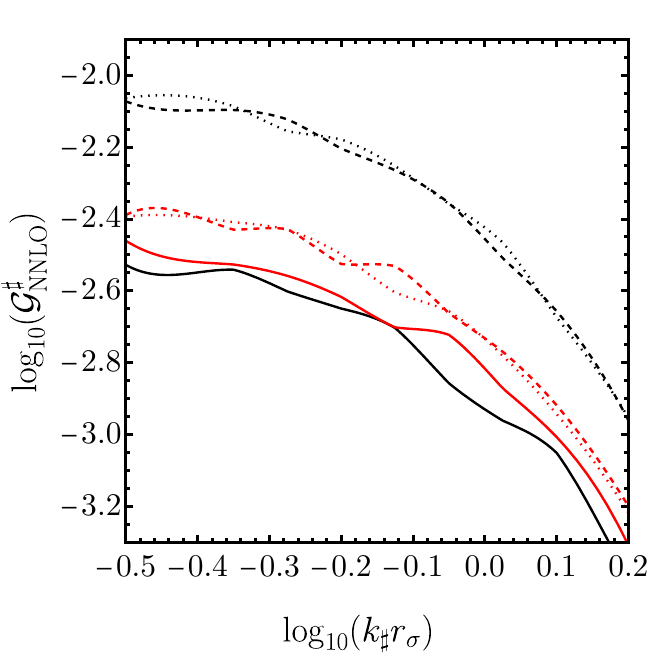}$$
\caption{\em \label{fig:gSharpNNLO} {
Numerical calculation of the function $\mathcal{G}^\sharp_{\rm NNLO}$ for $w=1$ (black) and $w=0.5$ (red) for $k_\sharp\eta_r=1$, $10$, and $100$ (solid, dashed, and dotted lines, respectively).
}}
\end{center}
\end{figure}

\bigskip
\noindent{\bf Case 2: NLO, sharp scalar spectrum (NLO$\sharp$)}

\begin{figure}[t]
\begin{center}
$$\includegraphics[width=.41\textwidth]{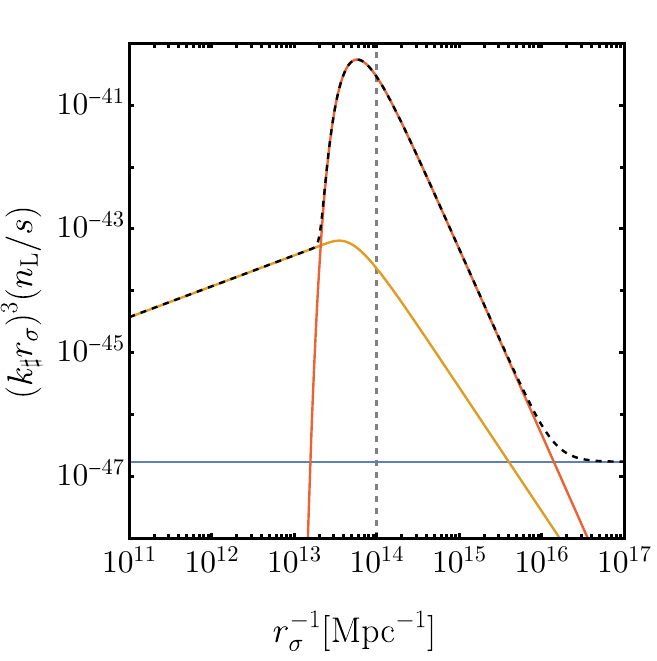}\qquad
\includegraphics[width=.5\textwidth]{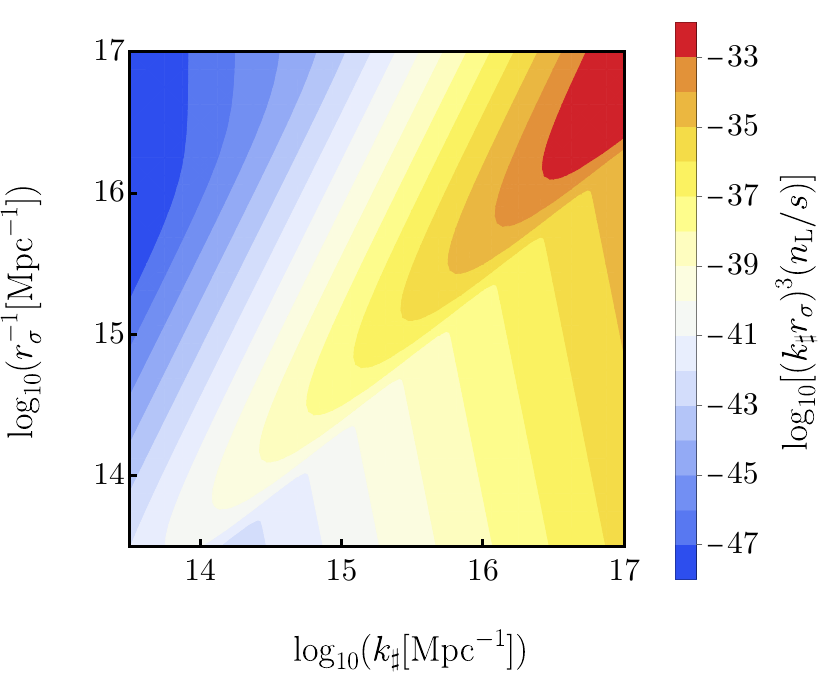}
$$
\caption{\em \label{fig:lepton_final} {
Left panel: sum of the ${\rm NLO}\flat$ (solid, blue), ${\rm NLO}\sharp$ (solid, orange), and ${\rm NNLO}\sharp$ (solid, red) contributions to the baryon asymmetry fluctuations, represented by the dashed black line and scaled by the factor $(k_\sharp r_\sigma)^3$. The plot is cut at the Silk scale $r_{\rm S}$, where the calculation stops being valid. We have taken $k_\sharp=10^{14}{\rm Mpc}^{-1}$ as a representative value for PBH dark matter (see Fig.\,\ref{fig:abu_mass}), together with $\mathcal{A}_\sharp=5\times 10^{-2}$, $r=0.01$, $\mathcal{A}_\flat=2\times 10^{-9}$ and $w=1$. Right panel: sum of the three contributions to the baryon asymmetry for the same parameters as in the left panel, but varying $k_\sharp$.
}}
\end{center}
\end{figure}

\noindent We now turn our attention to the cases in which the power spectrum is sharply peaked at some particular scale (not considered in \cite{Maroto:2022xrv}), and compute the corresponding enhancement to the baryon asymmetry fluctuations. The calculation in this case is exactly the same as the previous one up to the choice of the power spectra in eq.\,(\ref{eq:flat_spectra}). We now set
\begin{equation}
\mathcal{P}_\mathcal{R}(k)=\mathcal{A}_\sharp k_\sharp\delta(k-k_\sharp),\qquad \mathcal{P}_h(k)=r\mathcal{A}_\flat,
\end{equation}
and obtain
\begin{equation}
\frac{n_{\rm L}}{s}\bigg|_{\rm NLO\sharp}=\frac{45}{16\pi^4}\bigg(\frac{3+3w}{5+3w}\bigg)\frac{\sqrt{r\mathcal{A}_\flat\mathcal{A}_\sharp\mathcal{G}^\sharp_{\rm NLO}}}{g_{\star s}(T_0)(T_0r_\sigma)^3},
\end{equation}
with
\begin{align*}
\mathcal{G}^\sharp_{\rm NLO}&=\int_0^\infty dp_\sigma
\bigg|\int_0^\infty dx\; T_\phi\bigg(\frac{k_\sharp r_\sigma}{p_\sigma}x,k_\sharp r_\sigma\frac{\eta_r}{r_\sigma}\bigg)\frac{d}{dx}T_h\bigg(x,p_\sigma\frac{\eta_r}{r_\sigma}\bigg)\bigg|^2 \\
&\quad\exp\bigg[-\frac{1}{2}(p_\sigma^2+k_\sharp^2 r_\sigma^2)\bigg]\bigg[-3\frac{\cosh(p_\sigma k_\sharp r_\sigma)}{p_\sigma^3}+(3+p_\sigma^2k_\sharp^2r_\sigma^2)\frac{\sinh(p_\sigma k_\sharp r_\sigma)}{p_\sigma^4 k_\sharp r_\sigma}\bigg].\numberthis
\label{eq:nlo_sharp_fullint}
\end{align*}
Now the dimensionless factor $\mathcal{G}^\sharp_{\rm NLO}$ depends not only on $w$ and $\eta_r/r_\sigma$, but also on the parameter $k_\sharp r_\sigma$. It is once again instructive to obtain an analytical result using the approximations of the previous case. Setting $w=0$ and following the same procedure, we obtain
\begin{equation}
\label{eq:gsharp_approx}
\mathcal{G}^\sharp_{\rm NLO}=\frac{1}{k_\sharp r_\sigma}\sqrt{\frac{\pi}{2}}{\rm erf}\bigg(\frac{k_\sharp r_\sigma}{\sqrt{2}}\bigg)-e^{-k_\sharp^2 r_\sigma^2/2}\bigg(1+\frac{1}{3}k_\sharp^2r_\sigma^2\bigg).
\end{equation}
We remark that, as in the previous case, although we are interested in $w>1/3$, the choice $w=0$ is convenient because it allows us to obtain an analytical result. The numerical result for $w\neq 0$ is shown in Fig.\,\ref{fig:gsharp_approx}, where the dependence on $w$ can be seen to be very mild.

This function is highly peaked around $k_\sharp r_\sigma\simeq 3$, for which we find $\mathcal{G}^\sharp_{\rm NLO}\simeq 0.38$. This quantity has the asymptotic limits
\begin{equation}
\mathcal{G}^\sharp_{\rm NLO} \simeq
\begin{cases}
    \mathlarger{\frac{1}{15}}(k_\sharp r_\sigma)^4, & \text{for } k_\sharp r_\sigma \ll 1,
    \vspace{0.1in}
    \\
    \mathlarger{\frac{1}{k_\sharp r_\sigma}\sqrt{\frac{\pi}{2}}}, & \text{for } k_\sharp r_\sigma \gg 1.
\end{cases}
\label{eq:asymp_nlo}
\end{equation}
The function and its asymptotic limits are shown in the left panel of Fig.\,\ref{fig:gsharp_approx}. The numerical results for the full integral in eq.\,(\ref{eq:nlo_sharp_fullint}) are shown in the right panel of the same Figure for $k_\sharp r_\sigma=3$ and $k_\sharp r_\sigma=10$ as a function of $\eta_r/r_\sigma$ and $w$. We once again find that the dependence of the result on these two parameters is very mild, so only the scaling with $k_\sharp r_\sigma$ is relevant for our purposes. The dependence of the result on $r_\sigma$ is shown in Fig.\,\ref{fig:lepton_final} for values of $k_\sharp$ of interest for PBH dark matter, together with the NLO$\flat$ and NNLO$\sharp$ contributions, see the discussion below.

\bigskip
\noindent{\bf Case 3: NNLO, sharp scalar spectrum (NNLO$\sharp$)}

\noindent We now turn our attention to the NNLO contribution to the baryon asymmetry fluctuations. The amplitude of this term is entirely determined by the initial conditions and evolution of the scalar perturbations, and is therefore present even if the tensor-to-scalar ratio $r$ is vanishingly small. The calculation in this case is more involved than the previous ones, so we perform it in Appendix \ref{app:geom}. The resulting expression is
\begin{equation}
\label{eq:last_lepton}
\frac{n_{\rm L}}{s}\bigg|_{\rm NNLO\sharp}=\frac{45}{16\pi^4}\bigg(\frac{3+3w}{5+3w}\bigg)\frac{\sqrt{\mathcal{A}_\sharp^3\mathcal{G}^\sharp_{\rm NNLO}}}{g_{\star s}(T_0)}\bigg(\frac{k_\sharp}{T_0}\bigg)^3,
\end{equation}
where $\mathcal{G}^\sharp_{\rm NNLO}$ is computed in eq.\,(\ref{eq:gnnlo_app}). The coefficient $\mathcal{G}^\sharp_{\rm NNLO}$ is, in this case, a function of $w$ and $k_\sharp\eta_r$, as well as $k_\sharp r_\sigma$. The numerical calculation of this quantity is shown in Fig.\,\ref{fig:gSharpNNLO} for different values of the parameters. We once again find that only the dependence on $k_\sharp r_\sigma$ is relevant, and for $w=1$ and $k_\sharp\eta_r=1$ the function can be approximated by $\mathcal{G}^\sharp_{\rm NNLO}\simeq 10^{-2.8}\exp\big(-k_\sharp^2 r_\sigma^2\big)$. We adopt this expression for the following discussion.

The sum of the contributions in all three cases is shown on the left panel of Fig.\,\ref{fig:lepton_final} for the representative value $k_\sharp=10^{14}{\rm Mpc}^{-1}$, and on the right panel of the same Figure as a function of $k_\sharp$. We have rescaled $n_{\rm L}/s$ by the factor $(k_\sharp r_\sigma)^3$ to make the peaked structure of the spectrum more clear. This is the main result of this Section. We find that the NLO$\sharp$ contribution dominates for $r_\sigma^{-1}\ll k_\sharp$, and the NLO$\flat$ contribution becomes relevant only for $r_\sigma^{-1}\gg k_\sharp$. For $r_\sigma^{-1}\simeq k_\sharp$, the distribution is heavily peaked due to the NNLO$\sharp$ term, which yields an enhancement over the NLO$\flat$ result of $\mathcal{O}(10^6)$. As the right panel shows, the amplitude of the fluctuations at the peak of the distribution increases with $k_\sharp$. We note, however, that for $k_\sharp\simeq r_{\rm S}^{-1}$, where the enhancement is largest, the PBH masses obtained are far too small, as can be checked from eq.\,(\ref{eq:pbh_mass}), yielding a distribution of black holes that would have evaporated today \cite{Carr:2020gox} (though having a peak in the scalar spectrum at this scale is still possible, provided its amplitude is small enough so that PBHs are not overproduced). It is also clear from the Figure that the NLO$\sharp$ and NNLO$\sharp$ contributions do not change the distribution significantly on large scales, so the estimate in eq.\,(\ref{eq:hubble_est}) of the asymmetry on a Hubble-sized patch today holds in the presence of the new terms.

\phantomsection
\section*{Conclusions}
\addcontentsline{toc}{section}{Conclusions}

We have calculated the expressions for the mass and abundance of PBHs produced during an early epoch of reheating with a stiff equation of state. We find that the parameters that determine the black hole distribution are the scale at which the peak in the power spectrum is located $k_\sharp$, the temperature at which the transition to radiation occurs $T_r$, and the equation of state $w$. There are three relevant constraints in this scenario. The first is that, in order to reproduce the observed dark matter abundance, the black hole masses must be in the range (\ref{eq:mass_range}). The second is that for collapse to occur before the radiation era, the scale at which the peak in the spectrum is located must re-enter the horizon before the transition occurs, see eq.\,(\ref{eq:entry_con}). Finally, since tensor modes are enhanced in the presence of a stiff epoch, we must also ask that GWs are not overproduced so that the bound on their energy density today, which arises from BBN and CMB observations, is not violated, see eq.\,(\ref{eq:gw_con}). The allowed region of parameter space is shown in Fig.\,\ref{fig:abu_mass}.

In order to calculate the induced GW signal we have implemented a matching procedure for both the Green's function of the tensor modes and the transfer function of the scalar perturbations, thereby extending the results of \cite{Domenech:2019quo} by taking into account the full time evolution of these quantities. We confirm that, in the presence of a stiff epoch, the induced GW signal is enhanced. We have explicitly checked that the smoothness of the transition does not significantly change our results, and provided a procedure to compute the signal in these gradual decay scenarios exactly, without resorting to approximate analytical expressions of the Green's and transfer functions, extending the results of \cite{Inomata:2019ivs} to decaying fluids with stiff equations of state. As mentioned above, we have used these results to translate the bound on the GW abundance to a constraint on the PBH masses formed in this scenario. The results are shown in Figs.\,\ref{fig:abu_mass} and \ref{fig:gw_constraints}.

Finally, we have computed the chiral gravitational anomaly to third order in perturbations, and we find that the large scalar spectrum responsible for PBH formation induces a peak in the baryon asymmetry fluctuations on small scales. These results are shown in Fig.\,\ref{fig:lepton_final}. We have shown that this spectrum is essentially independent of the cosmological history (i.e.\,the reheating scale and equation of state) and is therefore a generic prediction present in every model of PBH formation from collapse induced by large density perturbations, even in the standard scenario where the black holes form during radiation domination, assuming only the matter content of the Standard Model. These fluctuations could, in principle, be used as an observable to probe not only the existence of PBHs, but also different models of inflation.  Assessing whether these fluctuations can be measured by any future experiments is beyond the scope of this paper, but we point out that the main obstacle to this end is the fact that they are much smaller than the observed background value, and therefore some sort of enhancement mechanism would likely be required in the particular scenario studied here. We remark, however, that the machinery we have developed could also be applied to other models of gravitational leptogenesis, such as the one in \cite{Alexander:2004us}, in which the fluctuations might be less suppressed.

The possibility of generating baryon isocurvature perturbations via the chiral gravitational anomaly was briefly mentioned in \cite{Maroto:2022xrv}. Since we remain completely agnostic about the inflationary model giving rise to the large curvature perturbations responsible for the peak in the power spectrum in our scenario, no definitive statements can be made about this possibility. We remark, however, that these fluctuations would only arise in multi-field models of inflation (or, potentially, in models in which the inflaton couples directly to $R\tilde{R}$, such as \cite{Alexander:2004us}). In the minimal single-field inflation scenario, the arguments in \cite{Weinberg:2003sw,Weinberg:2004kf,Weinberg:2004kr} apply, and isocurvature would be generated neither during inflation, nor during the subsequent reheating era.\footnote{In other words, the lepton asymmetry in our setup is only generated after inflation ends, as can be seen from the fact that the time integrals in $n_{\rm L}/s$ vanish at early times (see also Fig.\,2 of \cite{Maroto:2022xrv}), and thus inflationary perturbations are completely adiabatic.} The possibility of using different models of inflation to generate isocurvature would have to be studied on a case-by-case basis.

\section*{Acknowledgments}
We are grateful to Antonio Maroto and Alfredo Miravet for discussions about their work and the mechanism studied here, to Mehmet Gümüş for an extended discussion about the diagrams contributing to the six-point function entering the NNLO calculation, to Thomas Konstandin and Guillermo Ballesteros for comments on the draft, and to Alexander Westphal, Hyungjin Kim, and Mathias Pierre for discussions in the early stages of this project. This work is supported by the Deutsche Forschungsgemeinschaft under Germany's Excellence Strategy – EXC 2121 Quantum Universe – 390833306.

\appendix

\phantomsection
\section*{Appendices}
\addcontentsline{toc}{section}{Appendices}

\section{Evaluation of $\mathcal{G}^\sharp_{\rm NNLO}$}
\label{app:geom}

In this Appendix we compute the quantity $\mathcal{G}^\sharp_{\rm NNLO}$ relevant for the calculation of the lepton number density variance in the third case of Section \ref{sec:smooth_lepton}. By putting together eqs.\,(\ref{eq:nnlo_ori}) and (\ref{eq:smoothed_nl}) we obtain
\begin{align}
\big\langle a^6|n_{\rm L}|^2\big\rangle_{r_\sigma}&=\bigg(\frac{1}{32\pi^2}\bigg)^2\sum_{st}\int d\eta\int d\hat{\eta}\int\frac{d^3{\bm p}}{(2\pi)^3}\int\frac{d^3{\bm q}}{(2\pi)^3}\int\frac{d^3{\bm k}}{(2\pi)^3}\int\frac{d^3\hat{{\bm p}}}{(2\pi)^3}\int\frac{d^3\hat{{\bm q}}}{(2\pi)^3}\int\frac{d^3\hat{{\bm k}}}{(2\pi)^3}\nonumber \\
&\quad \bigg(\frac{3+3w}{5+3w}\bigg)^2T_\phi(q\eta)T_\phi(\hat{q}\hat{\eta})I_p'(\eta,k,|{\bm p}-{\bm k}|)I_{\hat{p}}'(\hat{\eta},\hat{k},|\hat{{\bm p}}-\hat{{\bm k}}|)\big\langle \mathcal{R}_{\bm{q}}\mathcal{R}_{\bm{k}}\mathcal{R}_{{\bm p}-{\bm k}} \mathcal{R}^\dagger_{\hat{\bm{q}}}\mathcal{R}^\dagger_{\hat{\bm{k}}}\mathcal{R}^\dagger_{\hat{{\bm p}}-\hat{{\bm k}}}\big\rangle\nonumber \\
&\quad \int d^3{\bm r}\int d^3\hat{{\bm r}}\;W_{r_\sigma}(r)W_{r_\sigma}(\hat{r}) e^{i({\bm p}+{\bm q})\cdot ({\bm x}+{\bm r})}e^{-i(\hat{{\bm p}}+\hat{{\bm q}})\cdot ({\bm x}+\hat{{\bm r}})} 
\nonumber \\
&\quad
\frac{1}{p^2\hat{p}^2}\Big[{\bm k}\cdot {\bm e}^s({\bm p})\cdot {\bm k}\Big]\Big[\hat{{\bm k}}\cdot {\bm e}^t(\hat{{\bm p}})\cdot \hat{{\bm k}}\Big]\Big[\epsilon_{ijk}\epsilon_{abc} q_j q_\ell p_k e_{i\ell}^s({\bm p})\hat{q}_b \hat{q}_d \hat{p}_c e_{ad}^t(\hat{{\bm p}})\Big].
\label{eq:hats}
\end{align}

We can evaluate the six-point function by using Wick's theorem
\begin{align}
\big[2(2\pi)^9\big]^{-1}\big\langle \mathcal{R}_{\bm{q}}\mathcal{R}_{\bm{k}}\mathcal{R}_{{\bm p}-{\bm k}} \mathcal{R}^\dagger_{\hat{\bm{q}}}\mathcal{R}^\dagger_{\hat{\bm{k}}}\mathcal{R}^\dagger_{\hat{{\bm p}}-\hat{{\bm k}}}\big\rangle
&=|\mathcal{R}_q|^2|\mathcal{R}_{|{\bm p}-{\bm k}|}|^2|\mathcal{R}_{\hat{q}}|^2
\delta^3_{{\bm q}+{\bm k}}\delta^3_{{\bm p}-{\bm k}-\hat{{\bm p}}+\hat{{\bm k}}}\delta^3_{\hat{{\bm q}}+\hat{{\bm k}}}\nonumber\\
&\,+|\mathcal{R}_q|^2|\mathcal{R}_k|^2|\mathcal{R}_{\hat{q}}|^2
\delta^3_{{\bm q}+{\bm p}-{\bm k}}\delta^3_{{\bm k}-\hat{{\bm k}}}\delta^3_{\hat{{\bm q}}+\hat{{\bm p}}-\hat{{\bm k}}}\nonumber\\
&\,+|\mathcal{R}_q|^2|\mathcal{R}_k|^2|\mathcal{R}_{\hat{q}}|^2
\delta^3_{{\bm q}-\hat{{\bm k}}}\delta^3_{{\bm k}-\hat{{\bm p}}+\hat{{\bm k}}}\delta^3_{\hat{{\bm q}}-{\bm p}+{\bm k}}\nonumber\\
&\,+|\mathcal{R}_q|^2|\mathcal{R}_{|{\bm p}-{\bm k}|}|^2|\mathcal{R}_{\hat{q}}|^2
\delta^3_{{\bm q}-\hat{{\bm p}}+\hat{{\bm k}}}\delta^3_{{\bm k}-\hat{{\bm q}}}\delta^3_{{\bm p}-{\bm k}-\hat{{\bm k}}},
\end{align}
where we have used the symmetry of the integrand under ${\bm k}\rightarrow {\bm p}-{\bm k}$ and $\hat{{\bm k}}\rightarrow \hat{{\bm p}}-\hat{{\bm k}}$ to simplify the result.\footnote{We have also ignored unphysical contact terms containing factors of the form $\delta^3_{\bm k}$, and one term with $\delta^3_{{\bm q}-\hat{{\bm q}}}\delta^3_{{\bm p}-\hat{{\bm p}}}$, which is not proportional to an overall momentum-conserving delta function $\delta^3_{{\bm p}+{\bm q}-\hat{{\bm p}}-\hat{{\bm q}}}$.} We can use the Dirac delta functions to perform the momentum integrals over ${\bm k}$, $\hat{{\bm k}}$ and $\hat{{\bm q}}$. After using the symmetry of $I_k$ under the exchange of the two momenta in the argument, we obtain
\begin{align}
\big\langle a^6|n_{\rm L}|^2\big\rangle_{r_\sigma}&=\frac{\pi^2}{32}\bigg(\frac{3+3w}{5+3w}\bigg)^2\sum_{st}\int d\eta\int d\hat{\eta}\int\frac{d^3{\bm p}}{(2\pi)^3}\int\frac{d^3{\bm q}}{(2\pi)^3}\int\frac{d^3{\bm k}}{(2\pi)^3}\;\frac{1}{p^2k^2}\big|\widehat{W}_{r_\sigma}(|{\bm p}+{\bm q}|)\big|^2\nonumber \\
&\quad T_\phi(q\eta)T_\phi(|{\bm p}+{\bm q}-{\bm k}|\hat{\eta})\Big[\epsilon_{ijk}q_j q_\ell p_k e_{i\ell}^s({\bm p})\Big]\Big[\epsilon_{abc} (p_b+q_b-k_b) (p_d+q_d) k_c e_{ad}^t({\bm k})\Big]\nonumber \\
&\quad\frac{\mathcal{P}_\mathcal{R}(q)}{q^3}\frac{\mathcal{P}_\mathcal{R}(|{\bm p}+{\bm q}-{\bm k}|)}{|{\bm p}+{\bm q}-{\bm k}|^3}\bigg\{
\frac{\mathcal{P}_\mathcal{R}(|{\bm p}+{\bm q}|)}{|{\bm p}+{\bm q}|^3}\Big[{\bm q}\cdot {\bm e}^s({\bm p})\cdot {\bm q}\Big]\Big[({\bm p}+{\bm q})\cdot {\bm e}^t({\bm k})\cdot ({\bm p}+{\bm q})\Big]\nonumber\\
&\quad I_p'(\eta,q,|{\bm p}+{\bm q}|)I_k'(\hat{\eta},|{\bm p}+{\bm q}-{\bm k}|,|{\bm p}+{\bm q}|)\nonumber\\
&\quad+
\frac{\mathcal{P}_\mathcal{R}(|{\bm q}-{\bm k}|)}{|{\bm q}-{\bm k}|^3}
\Big[({\bm q}-{\bm k})\cdot {\bm e}^s({\bm p})\cdot ({\bm q}-{\bm k})\Big]\Big[{\bm q}\cdot {\bm e}^t({\bm k})\cdot {\bm q}\Big]\nonumber\\
&\quad I_p'(\eta,|{\bm p}+{\bm q}-{\bm k}|,|{\bm q}-{\bm k}|)I_k'(\hat{\eta},q,|{\bm q}-{\bm k}|)\bigg\},
\label{eq:nnlo_long}
\end{align}
where we have also renamed the remaining dummy variable $\hat{{\bm p}}\rightarrow {\bm k}$. By choosing the sharp power spectrum in eq.\,(\ref{eq:pr_sharp}) and following the same procedure as for the two NLO cases in Section \ref{sec:smooth_lepton}, we find eq.\,(\ref{eq:last_lepton}), with
\begin{align}
\mathcal{G}^\sharp_{\rm NNLO}&\equiv\frac{2\pi^6}{k_\sharp^3}\int\frac{d^3{\bm p}}{(2\pi)^3}\int\frac{d^3{\bm q}}{(2\pi)^3}\int\frac{d^3{\bm k}}{(2\pi)^3}\frac{\delta(q-k_\sharp)}{k_\sharp^3}\frac{\delta(|{\bm p}+{\bm q}|-k_\sharp)}{k_\sharp^3}\frac{\delta(|{\bm p}+{\bm q}-{\bm k}|-k_\sharp)}{k_\sharp^3}\frac{1}{p^2k^2}\nonumber\\
&\quad \int d\eta\; T_\phi(k_\sharp\eta)I_p'(\eta,k_\sharp,k_\sharp)\int d\hat{\eta}\; T_\phi(k_\sharp\hat{\eta})I_k'(\hat{\eta},k_\sharp,k_\sharp)\nonumber\\
&\quad
\sum_{st}\Big[{\bm q}\cdot {\bm e}^s({\bm p})\cdot {\bm q}\Big]\Big[({\bm p}+{\bm q})\cdot {\bm e}^t({\bm k})\cdot ({\bm p}+{\bm q})\Big]\nonumber\\
&\quad\Big[F_{st}({\bm p},{\bm q},{\bm k})\big|\widehat{W}_{r_\sigma}(|{\bm p}+{\bm q}|)\big|^2+F_{ts}(-{\bm k},{\bm q},-{\bm p})\big|\widehat{W}_{r_\sigma}(|{\bm q}-{\bm k}|)\big|^2\Big],
\label{eq:gnnlo}
\end{align}
where
\begin{equation}
F_{st}({\bm p},{\bm q},{\bm k})\equiv \Big[\epsilon_{ijk}q_j q_\ell p_k e_{i\ell}^s({\bm p})\Big]\Big[\epsilon_{abc} (p_b+q_b-k_b) (p_d+q_d) k_c e_{ad}^t({\bm k})\Big].
\end{equation}
To obtain this expression from eq.\,(\ref{eq:nnlo_long}) we have renamed the dummy variables ${\bm k}\rightarrow -{\bm p}$ and ${\bm p}\rightarrow -{\bm k}$, as well as $s\leftrightarrow t$, in the second term inside the brackets.

Let us orient our coordinate system in such a way that ${\bm k}$ coincides with the ${\bm z}$ axis, and denote the angle between ${\bm p}$ and ${\bm q}$ by $\Theta$, and the angle between ${\bm p}+{\bm q}$ and ${\bm k}$ by $\Phi$. We can then use the following property of the Dirac delta function,
\begin{equation}
\delta[f(x)]=\sum_j\frac{\delta(x-x_j)}{|f'(x_j)|},
\end{equation}
where $x_j$ are the roots of $f(x)$, to find
\begin{align}
\delta(|{\bm p}+{\bm q}|-k_\sharp)&=\frac{1}{|\cos\Theta|}\delta(p+2k_\sharp\cos\Theta),\\
\delta(|{\bm p}+{\bm q}-{\bm k}|-k_\sharp)&=\frac{1}{|\cos\Phi|}\delta(k-2k_\sharp\cos\Phi).
\end{align}

We now switch to spherical coordinates in all of the momentum integrals in eq.\,(\ref{eq:gnnlo}). We denote the polar and azimuthal angles of ${\bm p}$ by $\theta_p$ and $\phi_p$, respectively, and similarly for ${\bm q}$ and ${\bm k}$. Since ${\bm k}$ coincides with the ${\bm z}$ axis, we can immediately perform the integrals over $\theta_k$ and $\phi_k$. We can also perform the integrals over the moduli by using the Dirac delta functions and including a Heaviside function for each to take into account the fact that the moduli must be positive. We find
\begin{align}
\mathcal{G}^\sharp_{\rm NNLO}&=\frac{1}{16\pi^2}\int d\theta_p \int  d\phi_p\int d\theta_q \int d\phi_q \sin\theta_p\sin\theta_q \Theta_{\rm H}(-\cos\Theta)\Theta_{\rm H}(\cos\Phi)\nonumber\\
&\quad \int dx\; T_\phi(x)\frac{d}{dx}I_{-2k_\sharp\cos\Theta}\bigg(\frac{x}{k_\sharp},k_\sharp,k_\sharp\bigg)\int d\hat{x}\; T_\phi(\hat{x})\frac{d}{d\hat{x}}I_{2k_\sharp\cos\Phi}\bigg(\frac{\hat{x}}{k_\sharp},k_\sharp,k_\sharp\bigg)\nonumber\\
&\quad
\sum_{st}\Big[\hat{{\bm q}}\cdot {\bm e}^s(\hat{{\bm p}})\cdot \hat{{\bm q}}\Big]\Big[(-2\cos\Theta\hat{{\bm p}}+\hat{{\bm q}})\cdot {\bm e}^t(\hat{{\bm k}})\cdot (-2\cos\Theta\hat{{\bm p}}+\hat{{\bm q}})\Big]\nonumber\\
&\quad\Big[\widehat{F}_{st}(\hat{{\bm p}},\hat{{\bm q}},\hat{{\bm k}})\big|\widehat{W}_{k_\sharp r_\sigma}(|\hat{{\bm q}}-2\cos\Theta\hat{{\bm p}}|)\big|^2
+\widehat{F}_{ts}(-\hat{{\bm k}},\hat{{\bm q}},-\hat{{\bm p}})\big|\widehat{W}_{k_\sharp r_\sigma}(|\hat{{\bm q}}-2\cos\Phi\hat{{\bm k}}|)\big|^2\Big],
\label{eq:gnnlo_app}
\end{align}
where $x\equiv k_\sharp\eta$, the hatted vectors are normalized to unity (and bear no relation to the hatted dummy variables in eq.\,(\ref{eq:hats})), and
\begin{align}
\widehat{F}_{st}(\hat{{\bm p}},\hat{{\bm q}},\hat{{\bm k}})&\equiv \Big[\epsilon_{ijk}\hat{q}_j \hat{q}_\ell \hat{p}_k e_{i\ell}^s(\hat{{\bm p}})\Big]\Big[\epsilon_{abc} \Big(\hat{q}_b-2\cos\Theta \hat{p}_b-2\cos\Phi \hat{k}_b\Big) \Big(\hat{q}_d-2\cos\Theta\hat{p}_d\Big) \hat{k}_c e_{ad}^t(\hat{{\bm k}})\Big],\nonumber\\
\widehat{F}_{ts}(-\hat{{\bm k}},\hat{{\bm q}},-\hat{{\bm p}})&\equiv \Big[\epsilon_{ijk}\hat{q}_j \hat{q}_\ell \hat{k}_k e_{i\ell}^t(\hat{{\bm k}})\Big]\Big[\epsilon_{abc} \Big(\hat{q}_b-2\cos\Theta\hat{p}_b-2\cos\Phi\hat{k}_b\Big) \Big(\hat{q}_d-2\cos\Phi\hat{k}_d\Big) \hat{p}_c e_{ad}^s(\hat{{\bm p}})\Big].
\label{eq:fhat}
\end{align}

The window functions can be explicitly written as
\begin{align}
\big|\widehat{W}_{k_\sharp r_\sigma}(|\hat{{\bm q}}-2\cos\Theta\hat{{\bm p}}|)\big|^2&=e^{-k_\sharp^2 r_\sigma^2/2},\nonumber\\
\big|\widehat{W}_{k_\sharp r_\sigma}(|\hat{{\bm q}}-2\cos\Phi\hat{{\bm k}}|)\big|^2&=e^{-k_\sharp^2 r_\sigma^2/2}\exp\Big[2k_\sharp^2 r_\sigma^2(\cos\theta_q-\cos\Phi)\cos\Phi\Big].
\end{align}
We can also write the $\Theta$ and $\Phi$ angles explicitly as follows,
\begin{align}
\cos\Theta&=\sin\theta_p\cos\phi_p\sin\theta_q\cos\phi_q+\sin\theta_p\sin\phi_p\sin\theta_q\sin\phi_q+\cos\theta_p\cos\theta_q,\\
\cos\Phi&=\cos\theta_q-2\cos\Theta\cos\theta_p.
\end{align}

We can write the expressions in eq.\,(\ref{eq:fhat}) explicitly by expanding the Levi-Civita symbols. After some straightforward algebra, we find
\begin{align}
\widehat{F}_{st}(\hat{\bm{p}},\hat{\bm{q}},\hat{\bm{k}})=
&-\Big[\hat{\bm q}\cdot(\hat{\bm q}-2\cos\Theta\hat{\bm p}-2\cos\Phi\hat{\bm k})\Big]\Big[\hat{\bm k}\cdot {\bm e}^s(\hat{\bm p})\cdot \hat{\bm q}\Big]\Big[\hat{\bm p}\cdot {\bm e}^t(\hat{\bm k})\cdot (\hat{\bm q}-2\cos\Theta\hat{\bm p})\Big]\nonumber\\
&+\Big[\hat{\bm p}\cdot (\hat{\bm q}-2\cos\Theta\hat{\bm p}-2\cos\Phi\hat{\bm k})\Big]\Big[\hat{\bm k} \cdot {\bm e}^s(\hat{\bm p})\cdot \hat{\bm q}\Big]
\Big[\hat{\bm q}\cdot {\bm e}^t(\hat{\bm k})\cdot (\hat{\bm q}-2\cos\Theta\hat{\bm p})\Big]\nonumber\\
&-\Big[\hat{\bm p}\cdot \hat{\bm k}\Big]\Big[(\hat{\bm q}-2\cos\Phi\hat{\bm k})\cdot  {\bm e}^s(\hat{\bm p})\cdot \hat{\bm q}\Big]
\Big[\hat{\bm q}\cdot {\bm e}^t(\hat{\bm k})\cdot (\hat{\bm q}-2\cos\Theta\hat{\bm p})\Big]\nonumber\\
&+\Big[\hat{\bm q}\cdot \hat{\bm k}\Big]\Big[(\hat{\bm q}-2\cos\Phi\hat{\bm k})\cdot {\bm e}^s(\hat{\bm p})\cdot \hat{\bm q}\Big]
\Big[\hat{\bm p}\cdot {\bm e}^t(\hat{\bm k})\cdot (\hat{\bm q}-2\cos\Theta\hat{\bm p})\Big]\nonumber\\
&-\Big[\hat{\bm q}\cdot \hat{\bm k}\Big]\Big[\hat{\bm p}\cdot (\hat{\bm q}-2\cos\Theta\hat{\bm p}-2\cos\Phi\hat{\bm k}) \Big]\Big[\hat{\bm q}\cdot  {\bm e}^s(\hat{\bm p})\cdot {\bm e}^t(\hat{\bm k})\cdot(\hat{\bm q}-2\cos\Theta\hat{\bm p})\Big]\nonumber\\
&+\Big[\hat{\bm p}\cdot \hat{\bm k}\Big]\Big[\hat{\bm q}\cdot (\hat{\bm q}-2\cos\Theta\hat{\bm p}-2\cos\Phi\hat{\bm k})\Big]\Big[ \hat{\bm q}\cdot  {\bm e}^s(\hat{\bm p})\cdot {\bm e}^t(\hat{\bm k})\cdot(\hat{\bm q}-2\cos\Theta\hat{\bm p})\Big],
\label{eq:fhat_final}
\end{align}
and, similarly,
\begin{align}
\widehat{F}_{ts}(-\hat{\bm{k}},\hat{\bm{q}},-\hat{\bm{p}})=
&-\Big[\hat{\bm q}\cdot(\hat{\bm q}-2\cos\Theta\hat{\bm p}-2\cos\Phi\hat{\bm k})\Big]\Big[\hat{\bm p}\cdot {\bm e}^t(\hat{\bm k})\cdot \hat{\bm q}\Big]\Big[\hat{\bm k}\cdot {\bm e}^s(\hat{\bm p})\cdot (\hat{\bm q}-2\cos\Phi\hat{\bm k})\Big]\nonumber\\
&+\Big[\hat{\bm k}\cdot (\hat{\bm q}-2\cos\Theta\hat{\bm p}-2\cos\Phi\hat{\bm k})\Big]\Big[\hat{\bm p} \cdot {\bm e}^t(\hat{\bm k})\cdot \hat{\bm q}\Big]
\Big[\hat{\bm q}\cdot {\bm e}^s(\hat{\bm p})\cdot (\hat{\bm q}-2\cos\Phi\hat{\bm k})\Big]\nonumber\\
&-\Big[\hat{\bm p}\cdot \hat{\bm k}\Big]\Big[(\hat{\bm q}-2\cos\Theta\hat{\bm p})\cdot  {\bm e}^t(\hat{\bm k})\cdot \hat{\bm q}\Big]
\Big[\hat{\bm q}\cdot {\bm e}^s(\hat{\bm p})\cdot (\hat{\bm q}-2\cos\Phi\hat{\bm k})\Big]\nonumber\\
&+\Big[\hat{\bm q}\cdot \hat{\bm p}\Big]\Big[(\hat{\bm q}-2\cos\Theta\hat{\bm p})\cdot {\bm e}^t(\hat{\bm k})\cdot \hat{\bm q}\Big]
\Big[\hat{\bm k}\cdot {\bm e}^s(\hat{\bm p})\cdot (\hat{\bm q}-2\cos\Phi\hat{\bm k})\Big]\nonumber\\
&-\Big[\hat{\bm q}\cdot \hat{\bm p}\Big]\Big[\hat{\bm k}\cdot (\hat{\bm q}-2\cos\Theta\hat{\bm p}-2\cos\Phi\hat{\bm k}) \Big]\Big[\hat{\bm q}\cdot  {\bm e}^t(\hat{\bm k})\cdot {\bm e}^s(\hat{\bm p})\cdot(\hat{\bm q}-2\cos\Phi\hat{\bm k})\Big]\nonumber\\
&+\Big[\hat{\bm p}\cdot \hat{\bm k}\Big]\Big[\hat{\bm q}\cdot (\hat{\bm q}-2\cos\Theta\hat{\bm p}-2\cos\Phi\hat{\bm k})\Big]\Big[ \hat{\bm q}\cdot  {\bm e}^t(\hat{\bm k})\cdot {\bm e}^s(\hat{\bm p})\cdot(\hat{\bm q}-2\cos\Phi\hat{\bm k})\Big].
\label{eq:fhat_flipped}
\end{align}

The ${\bm e}^s$ matrices are in general given by
\begin{equation}
e_{ij}^t(\hat{{\bm k}})=\frac{1}{\sqrt{2}}\delta^{t+}(e_ie_j-\bar{e}_i\bar{e}_j)+\frac{1}{\sqrt{2}}\delta^{t\times}(e_i\bar{e}_j+\bar{e}_ie_j),
\end{equation}
where $e_i$ and $\bar{e}_i$ are orthonormal vectors orthogonal to $\hat{{\bm k}}$. Since ${\bm k}$ coincides with the ${\bm z}$ axis, we can choose $e_i=x_i$ and $\bar{e}_i=y_i$ for ${\bm e}^t(\hat{\bm k})$. For ${\bm e}^s(\hat{\bm p})$ we can instead choose $e_i=v_i$ and $\bar{e}_i=w_i$, where
\begin{align}
v_i&=\cos\theta_p\cos\phi_p x_i+\cos\theta_p\sin\phi_py_i-\sin\theta_pz_i,\\
w_i&=-\sin\phi_px_i+\cos\phi_py_i.
\end{align}
These vectors can be easily checked to be orthogonal to $p$ by using \begin{equation}
p_i=p\sin\theta_p\cos\phi_p x_i+p\sin\theta_p\sin\phi_p y_i+p\cos\theta_p z_i.
\end{equation}
With these matrices in hand, the dot products in eqs.\,(\ref{eq:fhat_final}, \ref{eq:fhat_flipped}) can be written in terms of $\theta_p$, $\theta_q$, $\phi_p$, and $\phi_q$.

\begin{figure}[t]
\begin{center}
$$\includegraphics[width=.54\textwidth]{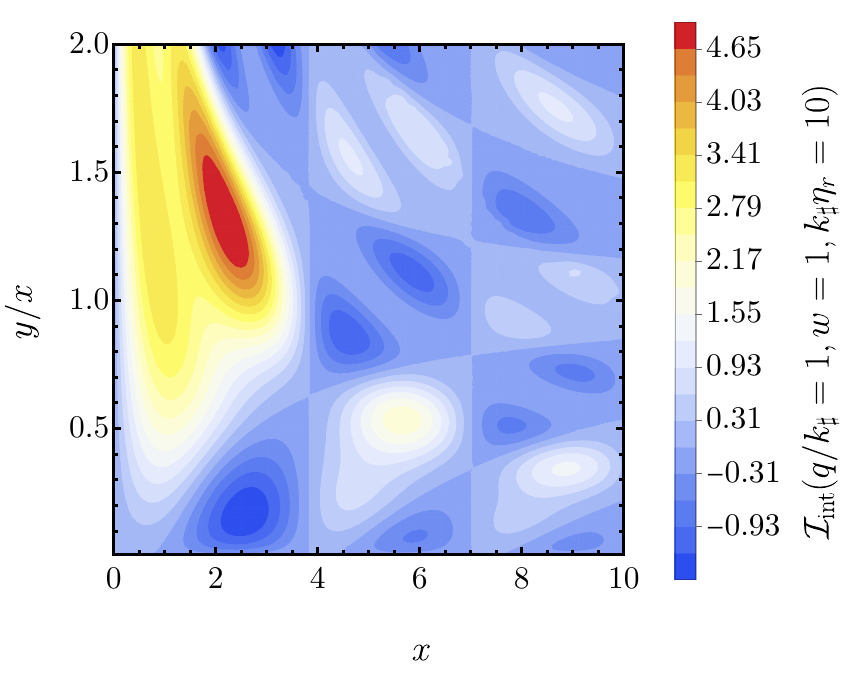}\qquad \includegraphics[width=.42\textwidth]{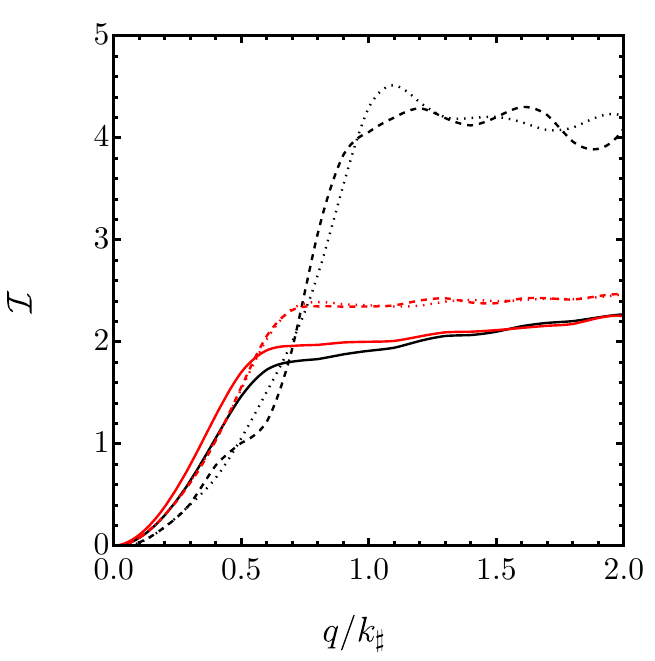}$$
\caption{\em \label{fig:qint_app} {
Left panel: integrand of $\mathcal{I}$ ($\mathcal{I}_{\rm int}$) for the specific parameters shown in the label. The integral converges quickly for $x\gtrsim 4$. Right panel: integral $\mathcal{I}$ as a function of $q/k_\sharp$ for $w=1$ (black) and $w=0.5$ (red) for $k_\sharp\eta_r=1$, $10$, and $100$ (solid, dashed, and dotted lines, respectively).
}}
\end{center}
\end{figure}

Let us turn our attention to the time integrals in eq.\,(\ref{eq:gnnlo_app}),
\begin{equation}
\mathcal{I}\equiv \int dx\; T_\phi(x)\frac{d}{dx}I_q\bigg(\frac{x}{k_\sharp},k_\sharp,k_\sharp\bigg).
\label{eq:iapp}
\end{equation}
The $I_q$ function is given by eq.\,(\ref{eq:i_function}), so that
\begin{equation}
\frac{d}{dx}I_q\bigg(\frac{x}{k_\sharp},k_\sharp,k_\sharp\bigg)=\bigg(\frac{3+3w}{5+3w}\bigg)^2\int_0^{x(q/k_\sharp)} \frac{d}{dx}qG_q\bigg(\frac{x}{k_\sharp},\frac{y}{q}\bigg)Q\bigg(\frac{y}{q},k_\sharp,k_\sharp\bigg) dy,
\end{equation}
with $y=q\eta'$, together with
\begin{align}
Q\bigg(\frac{y}{q},k_\sharp,k_\sharp\bigg)&=8T_\phi\bigg(\frac{k_\sharp}{q}y,k_\sharp\eta_r\bigg)^2+\frac{16}{3(1+p/\rho)}\bigg[T_\phi\bigg(\frac{k_\sharp}{q}y,k_\sharp\eta_r\bigg)+\frac{q}{\mathcal{H}}\frac{d}{dy}T_\phi\bigg(\frac{k_\sharp}{q}y,k_\sharp\eta_r\bigg)\bigg]^2,
\label{eq:qapp}
\end{align}
and
\begin{equation}
qG_q\bigg(\frac{x}{k_\sharp},\frac{y}{q}\bigg)=h_1\bigg(\frac{q}{k_\sharp}x,\frac{q}{k_\sharp}k_\sharp\eta_r\bigg)G_2\bigg(y,\frac{q}{k_\sharp}k_\sharp\eta_r\bigg)-h_2\bigg(\frac{q}{k_\sharp}x,\frac{q}{k_\sharp}k_\sharp\eta_r\bigg)G_1\bigg(y,\frac{q}{k_\sharp}k_\sharp\eta_r\bigg).
\end{equation}
In these equations we have highlighted the dependence of the transfer function $T_\phi$ on the parameter $k_\sharp\eta_r$. Similarly, the factor $q/\mathcal{H}$ in eq.\,(\ref{eq:qapp}) is a function of $y$ and $q\eta_r=(q/k_\sharp)k_\sharp\eta_r$. For the Green's function, we think of the independent solutions $h_{1,2}$ and the functions $G_{1,2}$ defined in eqs.\,(\ref{eq:gfunc1},\,\ref{eq:gfunc2}) as functions of $(q\eta,q\eta_r)$. We therefore find that $I_q$ is a function of $(x,q/k_\sharp,k_\sharp\eta_r,w)$.

Computing the 8-dimensional integral in eq.\,(\ref{eq:gnnlo_app}) numerically is difficult, even with Monte Carlo methods. The strategy we adopt is to first calculate the integral $\mathcal{I}$ for different choices of the parameters $w$ and $k_\sharp\eta_r$ as a function of $q/k_\sharp$. The results are shown on the right panel of Fig.\,\ref{fig:qint_app}. Since $q/k_\sharp$ lies between $0$ and $2$ in eq.\,(\ref{eq:gnnlo_app}), we restrict our calculation to this range. The integrand of $\mathcal{I}$ is shown on the left panel of the same Figure. With these integrals in hand, we can then calculate the remaining four integrals over the angles in eq.\,(\ref{eq:gnnlo_app}) numerically by varying the remaining parameter, $k_\sharp r_\sigma$. The results are shown in Fig.\,\ref{fig:gSharpNNLO}.


\printbibliography

\end{document}